\documentclass[aps,twocolumn,preprintnumbers,amsmath,amssymb,superscriptaddress,longbibliography]{revtex4-1} 
\usepackage{graphicx}
\usepackage{txfonts}
\usepackage{mathptmx}
\usepackage{braket}
\usepackage{color}
\usepackage[breaklinks,colorlinks,bookmarks=false,citecolor=blue,linkcolor=blue,urlcolor=blue]{hyperref}
\pdfmapfile{+txfonts.map}

\begin{document}

\preprint{\href{https://doi.org/10.1103/PhysRevB.102.081103}{Phys. Rev. B {\bfseries 102}, 081103(R) (2020)}}

\title{
Electronic localization in twisted bilayer MoS$_2$ with small rotation angle
}

\author{Somepalli \surname{Venkateswarlu} }
\affiliation{
CY Cergy Paris Universit\'e, CNRS, Laboratoire de Physique Th\'eorique et Mod\'elisation (UMR 8089), 95302 Cergy-Pontoise, France}
\author{Andreas \surname{Honecker}}
\affiliation{
CY Cergy Paris Universit\'e, CNRS, Laboratoire de Physique Th\'eorique et Mod\'elisation (UMR 8089), 95302 Cergy-Pontoise, France}
\author{Guy \surname{Trambly de Laissardi\`ere}}
\affiliation{
 CY Cergy Paris Universit\'e, CNRS, Laboratoire de Physique Th\'eorique et Mod\'elisation (UMR 8089), 95302 Cergy-Pontoise, France}

\date{May 26, 2020}

\begin{abstract}

Moir\'e patterns are known to confine electronic states in transition 
metal dichalcogenide bilayers, thus generalizing the notion of magic 
angles discovered in twisted bilayer graphene to semiconductors. Here, we 
present a revised Slater-Koster tight-binding model that facilitates the first 
reliable and systematic studies of such states in twisted bilayer MoS$_2$ 
for the whole range of rotation angles $\theta$. We show that isolated 
bands appear at low energy for $\theta \lesssim 5 - 6^\circ$. Moreover, 
these bands become ``flatbands'', characterized by a vanishing average 
velocity, for the smallest angles $\theta \lesssim 2^\circ$.

\end{abstract}


\maketitle


{\it Introduction.--}
Electronic correlations, {\it i.e.}, the Coulomb interactions between 
electrons, can give rise to exotic states of matter, with notable examples 
including Mott insulators \cite{Mott1949} and superconductors 
\cite{BCS57}. Some of the phenomena observed, e.g., in the so-called 
high-temperature superconductors \cite{Lee06} continue to pose puzzles 
despite of decades of research. The discovery of electronic localization 
by a moir\'e pattern in twisted bilayer graphene 
\cite{LopesdosSantos07,Trambly10,SuarezMorell10,Bistritzer11} allows the 
realization of such phenomena in intrinsically only weakly correlated 2D 
materials thanks to the emergence of flatbands at low energies that 
enhances the importance of interactions. Research in this field has been 
boosted by the experimental discovery of correlated insulators 
\cite{Cao18_correlated} and unconventional superconducting states 
\cite{Cao18_unconventional}. In recent years, the broad family of 
transition metal dichalcogenides (TMDs) 
\cite{Wang15_reviewTMD,Liu15_reviewTMD,Duong17_review}, which offers a 
wide variety of possible rotationally stacked bilayer systems, has also 
prompted numerous experimental 
\cite{vanderZande14,Liu14,Huang14,Huang16,Zhang17,Trainer17,Yeh16, 
Lin18,Pan18,zhang2019} and theoretical 
\cite{Roldan14b,Fang15,Cao15,Wang15,Constantinescu15,Tan16, 
Lu17,Naik18,Conte19,Maity19,Tang20,Wu20,Lu20,Xian2020,Pan20} studies to 
understand such confined moir\'e states in semiconductor materials. Many 
of these studies analyze the interlayer distances, the possible atomic 
relaxation, the transition from a direct band gap in the monolayer system 
to an indirect band gap in bilayer systems, and more generally the effect 
of interlayer coupling in those twisted 2D systems with various rotation 
angles $\theta$. At small values of $\theta$, the emergence of flatbands 
has been established \cite{Naik18} from first-principles density 
functional theory calculations in twisted bilayer MoS$_2$ (tb-MoS$_2$), 
and observed in a 3$^\circ$ twisted bilayer WSe$_2$ sample by using 
scanning tunneling spectroscopy \cite{zhang2019}. Recently, it has been 
shown numerically \cite{Lu20} that Lithium intercalation in tb-MoS$_2$ 
increases interlayer coupling and thus promotes flatbands around the gap. 
There is also experimental evidence that moir\'e patterns may give rise to 
confined states due to the mismatch of the lattice parameters in 
MoS$_2$-WSe$_2$ heterobilayers \cite{Pan18}.

\begin{figure}[t!]
\centering
\includegraphics[width=\columnwidth]{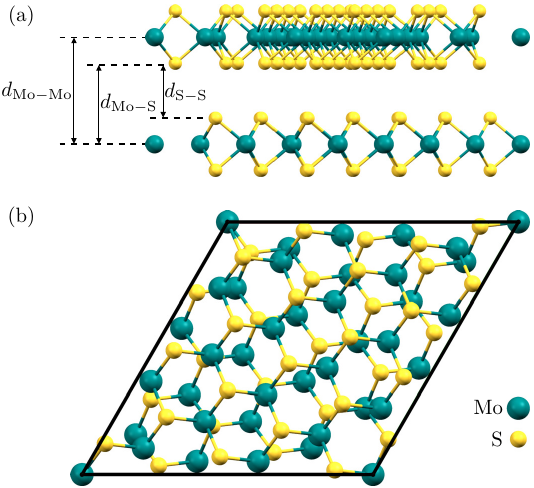}
\caption{
Atomic structure of bilayer MoS$_2$ at a twist
angle $\theta = 13.17^{\circ}$.
(a) side view.
(b) top view.} 
\label{Fig_structure2-3}
\end{figure}

Most theoretical investigations of the electronic structure of bilayer 
MoS$_2$ are density-functional theory (DFT) studies 
\cite{vanderZande14,Liu14,Huang14,Zhang17,Roldan14b,Fang15,Cao15,Wang15,Constantinescu15, 
Huang16,Tan16,Lu17,Fang15,Trainer17,Naik18,Debbichi14,Tao14,He14,sun2020effects} 
with eventually a Wannier wavefunction analysis \cite{Fang15}. Those 
approaches provide interesting results, but they do not allow a systematic 
analysis of the electronic structure as a function of the rotation angle 
$\theta$, in particular for small angles, {\it i.e.}, very large moir\'e 
cells, for which DFT calculations are not feasible. Several Tight-Binding 
(TB) models, based on Slater-Koster (SK) parameters \cite{Slater54}, have 
been proposed for monolayer MoS$_2$ 
\cite{Cappelluti13,Rostami13,Zahid13,Ridolfi15,SilvaGuillen16} and 
multi-layer MoS$_2$ \cite{Cappelluti13,Roldan14b,Fang15,Zahid13}. 
Following these efforts, we propose here a Slater-Koster set of parameters 
that match correctly the DFT bands around the gap of tb-MoS$_2$ with 
rotation angles $\theta > 7^\circ$. This SK-TB model, with the same 
parameters, is then used for smaller angles, in order to describe the 
confined moir\'e states. We thus show that, for $\theta \lesssim 6^\circ$, 
the valence band with the highest energy is separated from the other 
valence states by a minigap of a few meV. In addition, the width of this 
band decreases as $\theta$ decreases so that the average velocity of these 
electronic states reaches 0 for $\theta \lesssim 2^\circ$ such that flatbands emerge at these angles. This is reminiscent of the vanishing of the 
velocity at certain ``magic'' rotation angles in bilayer graphene 
\cite{LopesdosSantos07,Trambly10,SuarezMorell10,Bistritzer11,Trambly12} 
except that in the case of bilayer MoS$_2$ it arises for an interval of 
angles. Other minigaps and flatbands are also found in the conduction 
band. The confined states that are closest to the gap are localized in the 
AA stacking regions of the moir\'e pattern, like in twisted bilayer 
graphene.

{\it Atomic structure.--}
The commensurate structure of tb-MoS$_2$ can be defined in the same manner 
that is common for twisted bilayer graphene (see for instance 
Refs.~\cite{Campanera07,Mele10}). Here we use the same notation as in 
Refs.~\cite{Trambly10,Trambly12}. A commensurate tb-MoS$_2$ with rotation 
angle $\theta$ is defined by two integers $n$ and $m$, such that
\begin{equation}
\cos \theta = \frac{n^2+4nm+m^2}{2(n^2+nm+m^2)} ,
\end{equation} 
and its lattice vectors are $\vec{t}=n\vec{a}_1+m\vec{a}_2$ and 
$\vec{t{'}}=-m\vec{a}_1+(n+m)\vec{a}_2$, where $\vec{a}_1$ 
($a\sqrt{3}/2,-a/2,0)$ and $\vec{a}_2$ ($a\sqrt{3}/2,a/2,0)$ are the 
lattice vectors of monolayer MoS$_2$, with the lattice distance 
$a=0.318$\,nm. A unit cell of tb-MoS$_2$ contains $N=6(n^2+nm+m^2)$ atoms. 
Figure \ref{Fig_structure2-3} shows a ($n=2$, $m=3$) tb-MoS$_2$ unit cell 
containing 114 atoms. The cell $(\vec{a}_1,\vec{a}_2)$ of monolayer 
MoS$_2$ contains 3 atoms: Mo at $(0,0,0)$, S at $(0,a/\sqrt{3},0.49115a)$ 
and S at $(0,a/\sqrt{3},-0.49115a)$ \cite{Huisman71,Ridolfi15}. Note that in 
tb-MoS$_2$ different types of moir\'e cells can be built, as the atoms of 
a monolayer unit cell are not equivalent by symmetry (see Supplemental 
Material \cite{supplementaryMat}, section \ref{Sec_appendix_struc}). For 
simplicity, we consider only moir\'e patterns constructed as follows in 
the main text. Starting from an AA stacked bilayer (where Mo atoms of a 
layer lie above the Mo atoms of the other layer, and S atoms of a layer 
lie above the S atoms of the other layer), the layer 2 is rotated with 
respect to the layer 1 by the angle $\theta$ around a rotation axis going 
through two Mo atoms. We have checked that the qualitative results 
presented here are also found in tb-MoS$_2$ built from an AB stacked 
bilayer before rotation (see Supplemental Material 
\cite{supplementaryMat}). The interlayer distance between layers 
containing Mo atoms is fixed to $d_{\rm Mo-Mo} = 0.68$\,nm which is the 
DFT-optimized interlayer distance for AA stacked bilayer MoS$_2$. The 
atomic relaxation probably has an important effect on the electronic 
structure in tb-MoS$_2$ \cite{Naik18,Maity19}, like in twisted bilayer 
graphene \cite{Nam17}.  However in this work, our aim is to provide a 
simple tight-binding (TB) scheme using Slater-Koster parameters that can 
be used for tb-MoS$_2$ at all angles in order to analyze qualitatively the 
electronic states that are confined by the moir\'e pattern. Indeed, as was 
the case for twisted bilayer graphene, the study of the non-relaxed 
structure should make it possible to identify generic properties that will 
persist with relaxation. Therefore, our numerical results hould be 
qualitatively relevant even if they may not be quantitatively accurate.

\begin{figure}[t!]
\includegraphics[width=1.0\columnwidth,keepaspectratio]{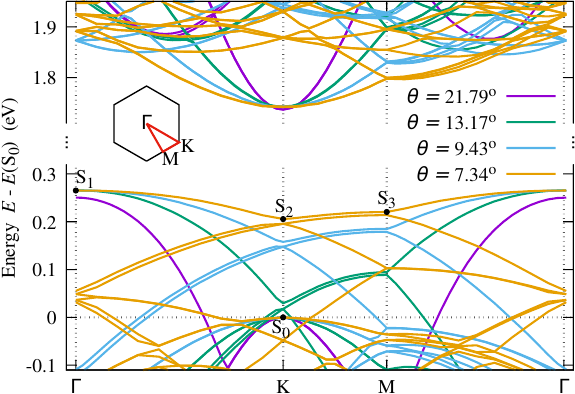}
\caption{\label{Fig_DFT_4angles}
DFT conduction and valence bands in tb-MoS$_2$: 
$(1,2)$ $\theta = 21.79^\circ$, 
$(2,3)$ $\theta = 13.17^\circ$, 
$(3,4)$ $\theta = 9.43^\circ$,
and $(4,5)$ $\theta = 7.34^\circ$. 
For every rotation angle, the origin of energy is fixed at the energy of 
the state S$_0$.
}
\end{figure}

{\it DFT calculations.--}
The DFT \cite{Hohenberg64,Kohn65} calculations were carried out with the 
ABINIT software \cite{Gonze02,Gonze09,Gonze16} within the Monkhorst-Pack 
scheme \cite{Zupan98} (more details are given in the Supplemental Material 
\cite{supplementaryMat}). LDA \cite{Jones89} and GGA + Van der Waals 
\cite{Perdew96} exchange-correlation functionals yield very similar 
results (see Fig.\ \ref{Fig_DFT_LDA_GGA-VdW} in the Supplemental Material 
\cite{supplementaryMat}), so all the results presented here are based on 
LDA calculations, which require less computation time for large systems.

Figure \ref{Fig_DFT_4angles} shows DFT bands of tb-MoS$_2$ along symmetric 
lines of the first Brillouin zone for four values of the rotation angle 
$\theta$. Note that the size of the Brillouin zones depends on the size of 
a unit cell of the moir\'e pattern such that the scale of the horizontal 
axis varies with $\theta$. These bands should be compared with the 
monolayer bands plotted in the same first Brillouin zone as follows. On 
the one hand, some bands are not affected by the value of $\theta$. 
Indeed, as shown in Fig.\ \ref{Fig_DFT_bi_mono} of the Supplemental 
Material \cite{supplementaryMat}, the parabolic band that emanates for the 
point S$_0$ is not affected by $\theta$. Therefore, we always set the 
energy of S$_0$ to zero. Similarly, for the angles shown in Fig.\ 
\ref{Fig_DFT_4angles}, the curvature of the parabola at the lowest 
conduction band energy at K is not affected by $\theta$. On the other 
hand, many bands are modified with respect to the monolayer case. Like for 
simple stacking bilayers (AA, AB, AB', \ldots) 
\cite{Debbichi14,Tao14,He14}, the highest valence energy at $\Gamma$, 
$E({\rm S}_1)$, increases with respect to the monolayer such that the gap 
becomes indirect. However, $E({\rm S}_1)$ does not vary significantly with 
$\theta$. In particular for the angles presented in Fig.\ 
\ref{Fig_DFT_4angles}, the curvature of the parabola at S$_1$ is not 
affected by $\theta$ and remains close to that of the monolayer.

Finally, considering the valence band, the most spectacular effect of 
decreasing $\theta$ is the increase of the energies of some bands, thus 
gradually filling the gap. This is, for instance, clearly seen in Fig.\ 
\ref{Fig_DFT_4angles} when considering the energy variation of the states 
S$_2$ and S$_3$ when $\theta$ decreases. Similarly, some energies of 
certain conduction bands decrease as $\theta$ decreases. Such a $\theta$ 
dependence of bands has already been observed for some values of the 
rotation angle in previous DFT calculations \cite{Naik18}. In order to 
analyze it systematically, it is necessary to perform calculation for 
smaller angles which is difficult using DFT calculations. This is the 
reason why we have developed a TB model that can be used for every value 
of $\theta$.

{\it TB calculations.--} 
In a first step, one needs to describe monolayer MoS$_2$ correctly. The 
states around the gap at the Fermi energy $E_F$ are mainly 4$d$ states of 
Mo \cite{Huisman71}. However, to describe valence and conduction bands 
correctly, it is not sufficient to restrict an effective Hamiltonian to 
4$d$ Mo orbitals. Indeed, the ligand field (S atoms) splits the 4$d$ 
levels of the transition metal (Mo) atoms, and thus creates a direct gap 
at the K point \cite{Huisman71}. Therefore, all TB models proposed in the 
literature include at least 3$p$ S orbitals 
\cite{Cappelluti13,Rostami13,Zahid13,Ridolfi15,SilvaGuillen16}. Roughly 
speaking the valence band has mainly $d_0 = 4d_{z^2}$ Mo character, 
whereas the conduction band has $d_0$ character mixed with $d_2 = 
4d_{x^2-y^2},\,4d_{xy}$ Mo character near the gap, and $d_1 = 4d_{xz},\, 
4d_{yz}$ Mo character for higher energies \cite{Cappelluti13}. It seems 
that $3p$ S orbitals, which have lower on-site energies, act as a 
perturbation of the $4d$ Mo bands. For this reason, several TB models 
\cite{Cappelluti13,Ridolfi15,SilvaGuillen16} fit rather well to the DFT 
band structure, while they propose very different parameters (on-site 
energies and Slater-Koster parameters). Our TB model for monolayer MoS$_2$ 
(Fig.\ \ref{Fig_mono_DFT_TB} in the Supplemental Material 
\cite{supplementaryMat}) is an adaptation of the model proposed in Ref.\ 
\cite{Ridolfi15} for monolayers, and is presented in detail in the 
Supplemental Material \cite{supplementaryMat}. Each unit cell of the 
monolayer contains 11 orbitals: 5 $d$ Mo orbitals ($d_0 = 4d_{z^2}$, $d_1 
= 4d_{xz},\, 4d_{yz}$, $d_2 = 4d_{x^2-y^2},\,4d_{xy}$ of 1 Mo atom) and 6 
$p$ S orbitals ($3p_x$, $3p_y$ and $3p_z$ of 2 S atoms). Since the precise 
model may differ for valence and conduction states \cite{Ridolfi15}, we 
have decided to focus on reproducing the valence band accurately. Note 
that our TB model has been adapted to simulate not only the DFT monolayer 
bands, but also the DFT bands of twisted bilayers (mainly valence bands), 
as shown in the following.

\begin{figure}[t!]
\begin{center}
\includegraphics[width=1.0\columnwidth,keepaspectratio]{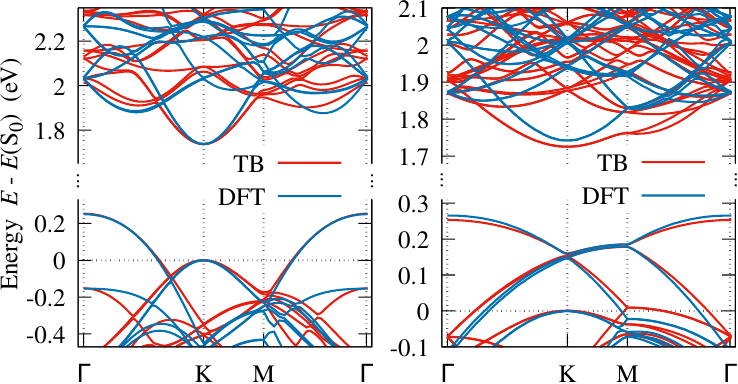}
\caption{\label{Fig_TB-DFT_bnds}
TB and DFT bands around the gap in tb-MoS$_2$: (left) $(1,2)$ $\theta = 21.79^\circ$ 
and (right) $(3,4)$ $\theta = 9.43^\circ$.
}
\end{center}
\end{figure}

In a second step, we consider the coupling between two layers of MoS$_2$. 
Most previous studies \cite{Cappelluti13,Roldan14b,Fang15,Zahid13} include 
only $p\,{\rm S}-p\,{\rm S}$ interlayer coupling terms, but $d\,{\rm 
Mo}-p\,{\rm S}$ and $d\,{\rm Mo}-d\,{\rm Mo}$ terms may also be important 
because we do not limit the interlayer coupling to first-neighbor hopping. 
Therefore, we include $p\,{\rm S}-p\,{\rm S}$, $d\,{\rm Mo}-p\,{\rm S}$, 
and $d\,{\rm Mo}-d\,{\rm Mo}$ interlayer terms in our Slater-Koster 
scheme. It turns out that the latter two are indeed important to reproduce 
the DFT valence band correctly. An exponential decay with inter-atomic 
distance \cite{Fang15} and a cutoff function \cite{Mehl96} are applied to 
of these interlayer terms, like in twisted bilayer graphene 
\cite{Trambly12}. Figure \ref{Fig_TB-DFT_bnds} shows the comparison 
between DFT and TB bands for tb-MoS$_2$ with $\theta = 21.79^\circ$ and 
$9.43^\circ$. The agreement is excellent for the highest energy valence 
bands and qualitatively correct for the conduction bands. All parameters 
of our TB model are given in the Supplemental Material 
\cite{supplementaryMat}.


\begin{figure}[t!]
\includegraphics[width=1.0\columnwidth,keepaspectratio]{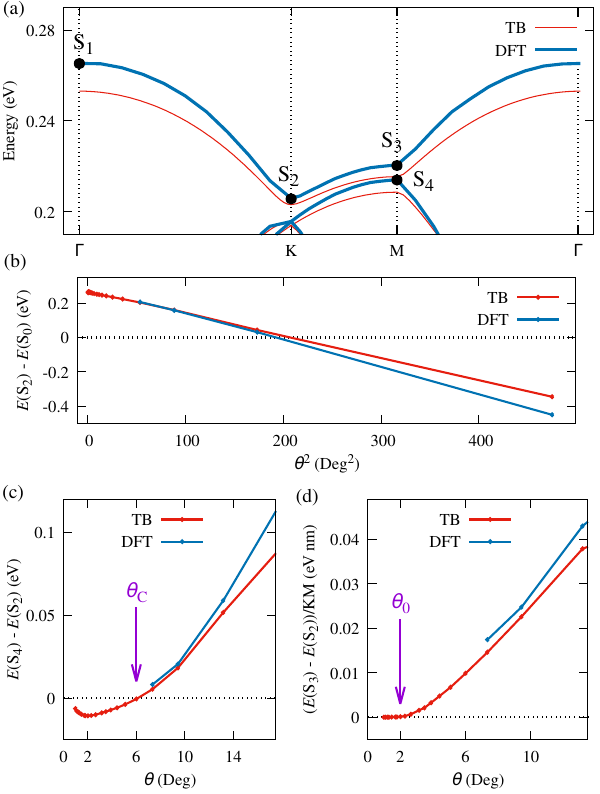}

\caption{\label{fig_analyse_Ei}
Dependence of valence bands on rotation angle $\theta$: (a) Valence band 
dispersion of $(4,5)$ tb-MoS$_2$, $\theta = 7.34^\circ$. (b) Energy 
$E({\rm S}_2)$ of the state S$_2$ (see panel (a)) versus $\theta^2$. (c) 
Energy difference between the states S$_4$ and S$_2$, $\Delta E_{24} = 
E({\rm S}_4) - E({\rm S}_2)$, versus $\theta$. A negative value of $\Delta 
E_{24}$ means that a gap $ | \Delta E_{24} | $ exists between the band 
below the gap and the other valence bands. (d) Average slope of $E(\vec 
k)$ of the band between states S$_2$ and S$_3$.
}
\end{figure}

{\it Effect of the rotation angle on bands.--} 
We now analyze the evolution of the bands around the main gap with 
$\theta$. Figure \ref{fig_analyse_Ei} shows this evolution for the top of 
the valence bands with a focus on the states labeled S$_1$, S$_2$, S$_3$, and 
S$_4$. Both DFT and TB results show that the energies $E({\rm S}_2)$ and 
$E({\rm S}_3)$ vary almost linearly with $\theta^2$ (Fig.\ 
\ref{fig_analyse_Ei}(b)), which is a strong indication that this 
phenomenon is a direct consequence of the moir\'e structure. Indeed, in 
the MoS$_2$ monolayer, the states around the gap are close to the $\Gamma$ 
and K points in reciprocal space, with a parabolic dispersion. In the 
twisted bilayer, the points $\Gamma_1$ and $\Gamma_2$ of the 2 monolayers 
(layer 1 and layer 2) coincide, while K$_1$ and K$_2$ are separated by a 
small distance ${\rm K}_1{\rm K}_2$ proportional to the angle $\theta$ for 
small $\theta$. As the monolayer band dispersion is parabolic, the energy 
of the crossing of the bands of the two layers varies with $\theta^2$, and 
so do the changes in energy induced by the moir\'e pattern. Similarly, 
many studies have shown that the changes of energy due to the moir\'e 
pattern in twisted graphene bilayer varies linearly with $\theta$ because 
the low-energy bands of a graphene monolayer are linear in 
$\|\vec{K}-\vec{k}\|$ (see, e.g., Ref.~\cite{Brihuega12}).

\begin{figure}[t!]
\begin{center}
\includegraphics[width=1.0\columnwidth,keepaspectratio]{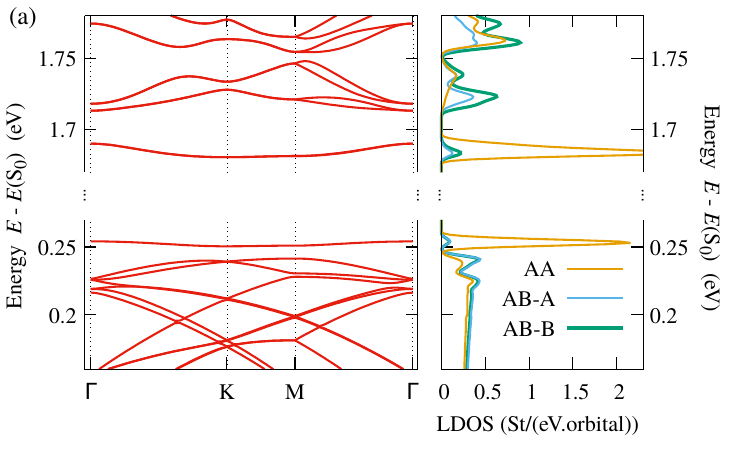}

\includegraphics[width=1.0\columnwidth,keepaspectratio]{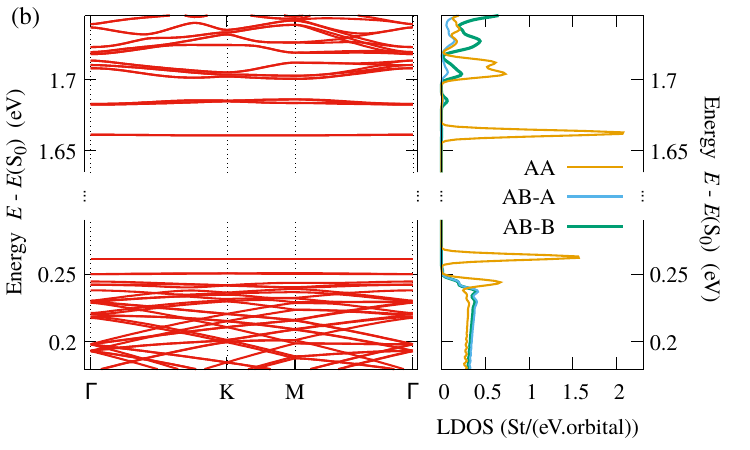}
\caption{\label{fig_LDOSAA_bnds}
TB band dispersion and local density of states (LDOS) of $d_0=d_{z^2}$ Mo 
atoms at the center of the AA stacking region and the center of the AB 
region: (a) for $(10,11)$ tb-MoS$_2$ $\theta = 3.15^\circ$, and (b) for 
$(20,21)$ tb-MoS$_2$ $\theta = 1.61^\circ$. In a moir\'e cell, two 
symmetrically equivalent AB stacking regions are located at $1/3$ and 
$2/3$ of the longest diagonal of the cell (see Sec.\ 
\ref{sec_built_fromAA} of the Supplemental Material 
\cite{supplementaryMat}). Each AB stacking region contains two types of Mo 
atoms: (AB-A) Mo atom of a layer lying above an S atom of the other layer, 
(AB-B) Mo atom of a layer not lying above an atom of the other layer.
}
\end{center}
\end{figure}

Furthermore, our TB computations show that the highest energy valence band 
is isolated from the remainder of the valence bands by a minigap for 
sufficiently small values of $\theta$ (Fig.\ \ref{fig_LDOSAA_bnds}). This 
is illustrated by Fig.\ \ref{fig_analyse_Ei}(c), showing that $E({\rm 
S}_4) - E({\rm S}_2) < 0$, {\it i.e.}, the presence of a minigap, for 
$\theta < \theta_C \approx 6^{\circ}$. This isolated band is not 
degenerate, thus it corresponds to one state per moir\'e cell. Figure 
\ref{fig_LDOSAA_bnds} shows that such isolated bands are also present 
among the conduction bands with different values of $\theta_C$. Finally, 
for the smallest angles, several isolated bands appear both among the 
valence and conduction bands.

We also consider the average slope of the highest valence band between the 
points K and M, {\it i.e.}, between the states S$_2$ and S$_3$ (Fig.\ 
\ref{fig_analyse_Ei}(a)). This quantity is proportional to the average 
Boltzmann velocity (intra-band velocity). As shown in Fig.\ 
\ref{fig_analyse_Ei}(d), this velocity tends towards zero for small 
angles, $\theta = \theta_0 \approx 2^\circ$. This demonstrates an 
electronic confinement corresponding to a ``flatband'', like it has been 
found for twisted bilayer graphene for specific angles, so-called magic 
angles 
\cite{LopesdosSantos07,Trambly10,SuarezMorell10,Bistritzer11,Trambly12}. 
However, in tb-MoS$_2$, this velocity vanishes not only for discrete 
values of $\theta$, but flatbands emerge for a continuous range of 
$\theta$, $\theta \lesssim \theta_0$.

{\it Confined state in the AA region of the moir\'e pattern.--}
Like for the monolayer, the electronic states of tb-MoS$_2$ closest to the 
gap have mainly $4d_{z^2}$ Mo character. This is still true for small 
angles, but states of the isolated bands are mainly localized in the AA 
stacking region. Consequently, the local density of states (LDOS) for 
$4d_{z^2}$ Mo at the center of AA region contains sharp peaks around the 
gap (Fig.\ \ref{fig_LDOSAA_bnds}). Note that in the LDOS (Fig.\ 
\ref{fig_LDOSAA_bnds}), the minigap discussed in the previous paragraph is 
not seen clearly because of the numerical Gaussian broadening used to 
calculate the LDOS. Other figures, presented in the Supplemental Material 
\cite{supplementaryMat}, show that the sharp peaks closest to the main gap 
are found neither in the LDOS of the other $4d$ Mo orbitals, nor in the 
LDOS of the Mo atoms that are not located in the AA stacking regions. 
Thus, the flatband states are confined in AA stacking regions, like in 
twisted bilayer graphene for small rotation angles 
\cite{Trambly10,Bistritzer11,Trambly12,LopesdosSantos12,Namarvar20}. The 
lowest-energy flatbands (closest to the gap in the valence and conduction 
bands) are localized at the center of the AA regions, as is also reflected 
by a strong enhancement of the local density of states in the 
corresponding regions (Fig.\ \ref{fig_LDOSAA_bnds}), whereas the next flatbands are localized in a ring in the AA regions rather than at the their 
center (see Fig.~\ref{Fig_20-21_VecP_G-K-M_ValenceB_average} in the 
Supplemental Material \cite{supplementaryMat}).

{\it Conclusion.--}
We have revisited the tight-binding description of twisted MoS$_2$ 
bilayers starting from DFT computations. Particular attention was paid to 
inerlayer Slater-Koster parameters and we confirmed that not only the 
closest $p\,{\rm S}-p\,{\rm S}$ interlayer coupling terms, but also 
$d\,{\rm Mo}-p\,{\rm S}$ and $d\,{\rm Mo}-d\,{\rm Mo}$ coupling needs to 
be taken into account for an accurate description. We then used this 
tight-binding model to investigate the band structure of MoS$_2$ bilayers 
at smaller rotation angles $\theta$ where the moir\'e unit cell becomes 
too large for DFT computations. We found that isolated bands appear in the 
valence and conduction bands close to the gap for $\theta \lesssim 5 - 
6^\circ$. For even smaller angles $\theta \lesssim 2^\circ$, the average 
velocity vanishes. The emergence of the corresponding flatbands is 
reflected by sharp peaks in the density of states. This phenomenon is 
accompanied by a localization of the wave function mainly in AA stacking 
regions. Depending on the flatband, this real-space confinement can occur 
at the center of AA region and also in a ring around the center of the AA 
region.

In the present discussion, we have focused on rotated MoS$_2$ bilayers 
that are constructed from AA stacking, but we have checked 
\cite{supplementaryMat} that qualitatively the same behavior is found when 
one starts from AB stacking instead.

The vanishing velocity and related emergence of flatbands identifies 
weakly doped MoS$_2$ bilayers as good candidates for the observation of 
strong correlation effects. Beyond first theoretical efforts in this 
direction \cite{Xian2020}, we offer our DFT-based tight-binding model as a 
solid starting point for more detailed studies of correlation effects in 
twisted MoS$_2$ bilayers.

Note added. Recently, we were alerted of two closely
related preprints \cite{Zhang2020a,Zhang2020b}.

{\it Acknowledgments.--}
The authors wish to thank L.\ Magaud, P.\ Mallet, D.\ Mayou, A.\ Missaoui, 
J.\ Vahedi, and J.-Y.\ Veuillen for fruitful discussions. Calculations 
have been performed at the Centre de Calculs (CDC), CY Cergy Paris 
Universit\'e and using HPC resources from GENCI-IDRIS (grant A0060910784). 
We thank Y.\ Costes and B.\ Mary, CDC, for computing assistance. This work 
was supported by the ANR project J2D (ANR-15-CE24-0017) and the 
Paris//Seine excellence initiative (grant 2017-231-C01-A0).


\bibliography{biblio}

\begin{thebibliography}{67}%
\makeatletter
\providecommand \@ifxundefined [1]{%
 \@ifx{#1\undefined}
}%
\providecommand \@ifnum [1]{%
 \ifnum #1\expandafter \@firstoftwo
 \else \expandafter \@secondoftwo
 \fi
}%
\providecommand \@ifx [1]{%
 \ifx #1\expandafter \@firstoftwo
 \else \expandafter \@secondoftwo
 \fi
}%
\providecommand \natexlab [1]{#1}%
\providecommand \enquote  [1]{``#1''}%
\providecommand \bibnamefont  [1]{#1}%
\providecommand \bibfnamefont [1]{#1}%
\providecommand \citenamefont [1]{#1}%
\providecommand \href@noop [0]{\@secondoftwo}%
\providecommand \href [0]{\begingroup \@sanitize@url \@href}%
\providecommand \@href[1]{\@@startlink{#1}\@@href}%
\providecommand \@@href[1]{\endgroup#1\@@endlink}%
\providecommand \@sanitize@url [0]{\catcode `\\12\catcode `\$12\catcode
  `\&12\catcode `\#12\catcode `\^12\catcode `\_12\catcode `\%12\relax}%
\providecommand \@@startlink[1]{}%
\providecommand \@@endlink[0]{}%
\providecommand \url  [0]{\begingroup\@sanitize@url \@url }%
\providecommand \@url [1]{\endgroup\@href {#1}{\urlprefix }}%
\providecommand \urlprefix  [0]{URL }%
\providecommand \Eprint [0]{\href }%
\providecommand \doibase [0]{http://dx.doi.org/}%
\providecommand \selectlanguage [0]{\@gobble}%
\providecommand \bibinfo  [0]{\@secondoftwo}%
\providecommand \bibfield  [0]{\@secondoftwo}%
\providecommand \translation [1]{[#1]}%
\providecommand \BibitemOpen [0]{}%
\providecommand \bibitemStop [0]{}%
\providecommand \bibitemNoStop [0]{.\EOS\space}%
\providecommand \EOS [0]{\spacefactor3000\relax}%
\providecommand \BibitemShut  [1]{\csname bibitem#1\endcsname}%
\let\auto@bib@innerbib\@empty
\bibitem [{\citenamefont {Mott}(1949)}]{Mott1949}%
  \BibitemOpen
  \bibfield  {author} {\bibinfo {author} {\bibfnamefont {N.~F.}\ \bibnamefont
  {Mott}},\ }\bibfield  {title} {\enquote {\bibinfo {title} {The basis of the
  electron theory of metals, with special reference to the transition
  metals},}\ }\href {\doibase 10.1088/0370-1298/62/7/303} {\bibfield  {journal}
  {\bibinfo  {journal} {Proc. Phys. Soc. A}\ }\textbf {\bibinfo {volume}
  {62}},\ \bibinfo {pages} {416--422} (\bibinfo {year} {1949})}\BibitemShut
  {NoStop}%
\bibitem [{\citenamefont {Bardeen}\ \emph {et~al.}(1957)\citenamefont
  {Bardeen}, \citenamefont {Cooper},\ and\ \citenamefont {Schrieffer}}]{BCS57}%
  \BibitemOpen
  \bibfield  {author} {\bibinfo {author} {\bibfnamefont {J.}~\bibnamefont
  {Bardeen}}, \bibinfo {author} {\bibfnamefont {L.~N.}\ \bibnamefont {Cooper}},
  \ and\ \bibinfo {author} {\bibfnamefont {J.~R.}\ \bibnamefont {Schrieffer}},\
  }\bibfield  {title} {\enquote {\bibinfo {title} {Theory of
  superconductivity},}\ }\href {\doibase 10.1103/PhysRev.108.1175} {\bibfield
  {journal} {\bibinfo  {journal} {Phys. Rev.}\ }\textbf {\bibinfo {volume}
  {108}},\ \bibinfo {pages} {1175--1204} (\bibinfo {year} {1957})}\BibitemShut
  {NoStop}%
\bibitem [{\citenamefont {Lee}\ \emph {et~al.}(2006)\citenamefont {Lee},
  \citenamefont {Nagaosa},\ and\ \citenamefont {Wen}}]{Lee06}%
  \BibitemOpen
  \bibfield  {author} {\bibinfo {author} {\bibfnamefont {P.~A.}\ \bibnamefont
  {Lee}}, \bibinfo {author} {\bibfnamefont {N.}~\bibnamefont {Nagaosa}}, \ and\
  \bibinfo {author} {\bibfnamefont {X.-G.}\ \bibnamefont {Wen}},\ }\bibfield
  {title} {\enquote {\bibinfo {title} {Doping a {M}ott insulator: Physics of
  high-temperature superconductivity},}\ }\href {\doibase
  10.1103/RevModPhys.78.17} {\bibfield  {journal} {\bibinfo  {journal} {Rev.
  Mod. Phys.}\ }\textbf {\bibinfo {volume} {78}},\ \bibinfo {pages} {17--85}
  (\bibinfo {year} {2006})}\BibitemShut {NoStop}%
\bibitem [{\citenamefont {Lopes~dos Santos}\ \emph {et~al.}(2007)\citenamefont
  {Lopes~dos Santos}, \citenamefont {Peres},\ and\ \citenamefont
  {Castro~Neto}}]{LopesdosSantos07}%
  \BibitemOpen
  \bibfield  {author} {\bibinfo {author} {\bibfnamefont {J.~M.~B.}\
  \bibnamefont {Lopes~dos Santos}}, \bibinfo {author} {\bibfnamefont
  {N.~M.~R.}\ \bibnamefont {Peres}}, \ and\ \bibinfo {author} {\bibfnamefont
  {A.~H.}\ \bibnamefont {Castro~Neto}},\ }\bibfield  {title} {\enquote
  {\bibinfo {title} {Graphene bilayer with a twist: Electronic structure},}\
  }\href {\doibase 10.1103/PhysRevLett.99.256802} {\bibfield  {journal}
  {\bibinfo  {journal} {Phys. Rev. Lett.}\ }\textbf {\bibinfo {volume} {99}},\
  \bibinfo {pages} {256802} (\bibinfo {year} {2007})}\BibitemShut {NoStop}%
\bibitem [{\citenamefont {Trambly~de Laissardi\`ere}\ \emph
  {et~al.}(2010)\citenamefont {Trambly~de Laissardi\`ere}, \citenamefont
  {Mayou},\ and\ \citenamefont {Magaud}}]{Trambly10}%
  \BibitemOpen
  \bibfield  {author} {\bibinfo {author} {\bibfnamefont {G.}~\bibnamefont
  {Trambly~de Laissardi\`ere}}, \bibinfo {author} {\bibfnamefont
  {D.}~\bibnamefont {Mayou}}, \ and\ \bibinfo {author} {\bibfnamefont
  {L.}~\bibnamefont {Magaud}},\ }\bibfield  {title} {\enquote {\bibinfo {title}
  {Localization of {D}irac electrons in rotated graphene bilayers},}\ }\href
  {\doibase 10.1021/nl902948m} {\bibfield  {journal} {\bibinfo  {journal} {Nano
  Letters}\ }\textbf {\bibinfo {volume} {10}},\ \bibinfo {pages} {804--808}
  (\bibinfo {year} {2010})}\BibitemShut {NoStop}%
\bibitem [{\citenamefont {Su\'arez~Morell}\ \emph {et~al.}(2010)\citenamefont
  {Su\'arez~Morell}, \citenamefont {Correa}, \citenamefont {Vargas},
  \citenamefont {Pacheco},\ and\ \citenamefont {Barticevic}}]{SuarezMorell10}%
  \BibitemOpen
  \bibfield  {author} {\bibinfo {author} {\bibfnamefont {E.}~\bibnamefont
  {Su\'arez~Morell}}, \bibinfo {author} {\bibfnamefont {J.~D.}\ \bibnamefont
  {Correa}}, \bibinfo {author} {\bibfnamefont {P.}~\bibnamefont {Vargas}},
  \bibinfo {author} {\bibfnamefont {M.}~\bibnamefont {Pacheco}}, \ and\
  \bibinfo {author} {\bibfnamefont {Z.}~\bibnamefont {Barticevic}},\ }\bibfield
   {title} {\enquote {\bibinfo {title} {Flat bands in slightly twisted bilayer
  graphene: Tight-binding calculations},}\ }\href {\doibase
  10.1103/PhysRevB.82.121407} {\bibfield  {journal} {\bibinfo  {journal} {Phys.
  Rev. B}\ }\textbf {\bibinfo {volume} {82}},\ \bibinfo {pages} {121407(R)}
  (\bibinfo {year} {2010})}\BibitemShut {NoStop}%
\bibitem [{\citenamefont {Bistritzer}\ and\ \citenamefont
  {MacDonald}(2011)}]{Bistritzer11}%
  \BibitemOpen
  \bibfield  {author} {\bibinfo {author} {\bibfnamefont {R.}~\bibnamefont
  {Bistritzer}}\ and\ \bibinfo {author} {\bibfnamefont {A.~H.}\ \bibnamefont
  {MacDonald}},\ }\bibfield  {title} {\enquote {\bibinfo {title} {Moir{\'e}
  bands in twisted double-layer graphene},}\ }\href {\doibase
  10.1073/pnas.1108174108} {\bibfield  {journal} {\bibinfo  {journal}
  {Proceedings of the National Academy of Sciences}\ }\textbf {\bibinfo
  {volume} {108}},\ \bibinfo {pages} {12233--12237} (\bibinfo {year}
  {2011})}\BibitemShut {NoStop}%
\bibitem [{\citenamefont {Cao}\ \emph {et~al.}(2018{\natexlab{a}})\citenamefont
  {Cao}, \citenamefont {Fatemi}, \citenamefont {Demir}, \citenamefont {Fang},
  \citenamefont {Tomarken}, \citenamefont {Luo}, \citenamefont
  {Sanchez-Yamagishi}, \citenamefont {Watanabe}, \citenamefont {Taniguchi},
  \citenamefont {Kaxiras}, \citenamefont {Ashoori},\ and\ \citenamefont
  {Jarillo-Herrero}}]{Cao18_correlated}%
  \BibitemOpen
  \bibfield  {author} {\bibinfo {author} {\bibfnamefont {Y.}~\bibnamefont
  {Cao}}, \bibinfo {author} {\bibfnamefont {V.}~\bibnamefont {Fatemi}},
  \bibinfo {author} {\bibfnamefont {A.}~\bibnamefont {Demir}}, \bibinfo
  {author} {\bibfnamefont {S.}~\bibnamefont {Fang}}, \bibinfo {author}
  {\bibfnamefont {S.~L.}\ \bibnamefont {Tomarken}}, \bibinfo {author}
  {\bibfnamefont {J.~Y.}\ \bibnamefont {Luo}}, \bibinfo {author} {\bibfnamefont
  {J.~D.}\ \bibnamefont {Sanchez-Yamagishi}}, \bibinfo {author} {\bibfnamefont
  {K.}~\bibnamefont {Watanabe}}, \bibinfo {author} {\bibfnamefont
  {T.}~\bibnamefont {Taniguchi}}, \bibinfo {author} {\bibfnamefont
  {E.}~\bibnamefont {Kaxiras}}, \bibinfo {author} {\bibfnamefont {R.~C.}\
  \bibnamefont {Ashoori}}, \ and\ \bibinfo {author} {\bibfnamefont
  {P.}~\bibnamefont {Jarillo-Herrero}},\ }\bibfield  {title} {\enquote
  {\bibinfo {title} {Correlated insulator behaviour at half-filling in
  magic-angle graphene superlattices},}\ }\href {\doibase 10.1038/nature26154}
  {\bibfield  {journal} {\bibinfo  {journal} {Nature}\ }\textbf {\bibinfo
  {volume} {556}},\ \bibinfo {pages} {80} (\bibinfo {year}
  {2018}{\natexlab{a}})}\BibitemShut {NoStop}%
\bibitem [{\citenamefont {Cao}\ \emph {et~al.}(2018{\natexlab{b}})\citenamefont
  {Cao}, \citenamefont {Fatemi}, \citenamefont {Fang}, \citenamefont
  {Watanabe}, \citenamefont {Taniguchi}, \citenamefont {Kaxiras},\ and\
  \citenamefont {Jarillo-Herrero}}]{Cao18_unconventional}%
  \BibitemOpen
  \bibfield  {author} {\bibinfo {author} {\bibfnamefont {Y.}~\bibnamefont
  {Cao}}, \bibinfo {author} {\bibfnamefont {V.}~\bibnamefont {Fatemi}},
  \bibinfo {author} {\bibfnamefont {S.}~\bibnamefont {Fang}}, \bibinfo {author}
  {\bibfnamefont {K.}~\bibnamefont {Watanabe}}, \bibinfo {author}
  {\bibfnamefont {T.}~\bibnamefont {Taniguchi}}, \bibinfo {author}
  {\bibfnamefont {E.}~\bibnamefont {Kaxiras}}, \ and\ \bibinfo {author}
  {\bibfnamefont {P.}~\bibnamefont {Jarillo-Herrero}},\ }\bibfield  {title}
  {\enquote {\bibinfo {title} {Unconventional superconductivity in magic-angle
  graphene superlattices},}\ }\href {\doibase 10.1038/nature26160} {\bibfield
  {journal} {\bibinfo  {journal} {Nature}\ }\textbf {\bibinfo {volume} {556}},\
  \bibinfo {pages} {43} (\bibinfo {year} {2018}{\natexlab{b}})}\BibitemShut
  {NoStop}%
\bibitem [{\citenamefont {Wang}\ \emph
  {et~al.}(2015{\natexlab{a}})\citenamefont {Wang}, \citenamefont {Yuan},
  \citenamefont {Sae~Hong}, \citenamefont {Li},\ and\ \citenamefont
  {Cui}}]{Wang15_reviewTMD}%
  \BibitemOpen
  \bibfield  {author} {\bibinfo {author} {\bibfnamefont {H.}~\bibnamefont
  {Wang}}, \bibinfo {author} {\bibfnamefont {H.}~\bibnamefont {Yuan}}, \bibinfo
  {author} {\bibfnamefont {S.}~\bibnamefont {Sae~Hong}}, \bibinfo {author}
  {\bibfnamefont {Y.}~\bibnamefont {Li}}, \ and\ \bibinfo {author}
  {\bibfnamefont {Y.}~\bibnamefont {Cui}},\ }\bibfield  {title} {\enquote
  {\bibinfo {title} {Physical and chemical tuning of two-dimensional transition
  metal dichalcogenides},}\ }\href {\doibase 10.1039/C4CS00287C} {\bibfield
  {journal} {\bibinfo  {journal} {Chem. Soc. Rev.}\ }\textbf {\bibinfo {volume}
  {44}},\ \bibinfo {pages} {2664--2680} (\bibinfo {year}
  {2015}{\natexlab{a}})}\BibitemShut {NoStop}%
\bibitem [{\citenamefont {Liu}\ \emph {et~al.}(2015)\citenamefont {Liu},
  \citenamefont {Xiao}, \citenamefont {Yao}, \citenamefont {Xu},\ and\
  \citenamefont {Yao}}]{Liu15_reviewTMD}%
  \BibitemOpen
  \bibfield  {author} {\bibinfo {author} {\bibfnamefont {G.-B.}\ \bibnamefont
  {Liu}}, \bibinfo {author} {\bibfnamefont {D.}~\bibnamefont {Xiao}}, \bibinfo
  {author} {\bibfnamefont {Y.}~\bibnamefont {Yao}}, \bibinfo {author}
  {\bibfnamefont {X.}~\bibnamefont {Xu}}, \ and\ \bibinfo {author}
  {\bibfnamefont {W.}~\bibnamefont {Yao}},\ }\bibfield  {title} {\enquote
  {\bibinfo {title} {Electronic structures and theoretical modelling of
  two-dimensional group-{VIB} transition metal dichalcogenides},}\ }\href
  {\doibase 10.1039/C4CS00301B} {\bibfield  {journal} {\bibinfo  {journal}
  {Chem. Soc. Rev.}\ }\textbf {\bibinfo {volume} {44}},\ \bibinfo {pages}
  {2643--2663} (\bibinfo {year} {2015})}\BibitemShut {NoStop}%
\bibitem [{\citenamefont {Duong}\ \emph {et~al.}(2017)\citenamefont {Duong},
  \citenamefont {Yun},\ and\ \citenamefont {Lee}}]{Duong17_review}%
  \BibitemOpen
  \bibfield  {author} {\bibinfo {author} {\bibfnamefont {D.~L.}\ \bibnamefont
  {Duong}}, \bibinfo {author} {\bibfnamefont {S.~J.}\ \bibnamefont {Yun}}, \
  and\ \bibinfo {author} {\bibfnamefont {Y.~H.}\ \bibnamefont {Lee}},\
  }\bibfield  {title} {\enquote {\bibinfo {title} {van der {W}aals layered
  materials: Opportunities and challenges},}\ }\href {\doibase
  10.1021/acsnano.7b07436} {\bibfield  {journal} {\bibinfo  {journal} {ACS
  Nano}\ }\textbf {\bibinfo {volume} {11}},\ \bibinfo {pages} {11803--11830}
  (\bibinfo {year} {2017})}\BibitemShut {NoStop}%
\bibitem [{\citenamefont {van~der Zande}\ \emph {et~al.}(2014)\citenamefont
  {van~der Zande}, \citenamefont {Kunstmann}, \citenamefont {Chernikov},
  \citenamefont {Chenet}, \citenamefont {You}, \citenamefont {Zhang},
  \citenamefont {Huang}, \citenamefont {Berkelbach}, \citenamefont {Wang},
  \citenamefont {Zhang}, \citenamefont {Hybertsen}, \citenamefont {Muller},
  \citenamefont {Reichman}, \citenamefont {Heinz},\ and\ \citenamefont
  {Hone}}]{vanderZande14}%
  \BibitemOpen
  \bibfield  {author} {\bibinfo {author} {\bibfnamefont {A.~M.}\ \bibnamefont
  {van~der Zande}}, \bibinfo {author} {\bibfnamefont {J.}~\bibnamefont
  {Kunstmann}}, \bibinfo {author} {\bibfnamefont {A.}~\bibnamefont
  {Chernikov}}, \bibinfo {author} {\bibfnamefont {D.~A.}\ \bibnamefont
  {Chenet}}, \bibinfo {author} {\bibfnamefont {Y.}~\bibnamefont {You}},
  \bibinfo {author} {\bibfnamefont {X.}~\bibnamefont {Zhang}}, \bibinfo
  {author} {\bibfnamefont {P.~Y.}\ \bibnamefont {Huang}}, \bibinfo {author}
  {\bibfnamefont {T.~C.}\ \bibnamefont {Berkelbach}}, \bibinfo {author}
  {\bibfnamefont {L.}~\bibnamefont {Wang}}, \bibinfo {author} {\bibfnamefont
  {F.}~\bibnamefont {Zhang}}, \bibinfo {author} {\bibfnamefont {M.~S.}\
  \bibnamefont {Hybertsen}}, \bibinfo {author} {\bibfnamefont {D.~A.}\
  \bibnamefont {Muller}}, \bibinfo {author} {\bibfnamefont {D.~R.}\
  \bibnamefont {Reichman}}, \bibinfo {author} {\bibfnamefont {T.~F.}\
  \bibnamefont {Heinz}}, \ and\ \bibinfo {author} {\bibfnamefont {J.~C.}\
  \bibnamefont {Hone}},\ }\bibfield  {title} {\enquote {\bibinfo {title}
  {Tailoring the electronic structure in bilayer molybdenum disulfide via
  interlayer twist},}\ }\href {\doibase 10.1021/nl501077m} {\bibfield
  {journal} {\bibinfo  {journal} {Nano Letters}\ }\textbf {\bibinfo {volume}
  {14}},\ \bibinfo {pages} {3869--3875} (\bibinfo {year} {2014})}\BibitemShut
  {NoStop}%
\bibitem [{\citenamefont {Liu}\ \emph {et~al.}(2014)\citenamefont {Liu},
  \citenamefont {Zhang}, \citenamefont {Cao}, \citenamefont {Jin},
  \citenamefont {Qiu}, \citenamefont {Zhou}, \citenamefont {Zettl},
  \citenamefont {Yang}, \citenamefont {Louie},\ and\ \citenamefont
  {Wang}}]{Liu14}%
  \BibitemOpen
  \bibfield  {author} {\bibinfo {author} {\bibfnamefont {K.}~\bibnamefont
  {Liu}}, \bibinfo {author} {\bibfnamefont {L.}~\bibnamefont {Zhang}}, \bibinfo
  {author} {\bibfnamefont {T.}~\bibnamefont {Cao}}, \bibinfo {author}
  {\bibfnamefont {C.}~\bibnamefont {Jin}}, \bibinfo {author} {\bibfnamefont
  {D.}~\bibnamefont {Qiu}}, \bibinfo {author} {\bibfnamefont {Q.}~\bibnamefont
  {Zhou}}, \bibinfo {author} {\bibfnamefont {A.}~\bibnamefont {Zettl}},
  \bibinfo {author} {\bibfnamefont {P.}~\bibnamefont {Yang}}, \bibinfo {author}
  {\bibfnamefont {S.~G.}\ \bibnamefont {Louie}}, \ and\ \bibinfo {author}
  {\bibfnamefont {F.}~\bibnamefont {Wang}},\ }\bibfield  {title} {\enquote
  {\bibinfo {title} {Evolution of interlayer coupling in twisted molybdenum
  disulfide bilayers},}\ }\href {\doibase 10.1038/ncomms5966} {\bibfield
  {journal} {\bibinfo  {journal} {Nat. Commun.}\ }\textbf {\bibinfo {volume}
  {5}},\ \bibinfo {pages} {4966} (\bibinfo {year} {2014})}\BibitemShut
  {NoStop}%
\bibitem [{\citenamefont {Huang}\ \emph {et~al.}(2014)\citenamefont {Huang},
  \citenamefont {Ling}, \citenamefont {Liang}, \citenamefont {Kong},
  \citenamefont {Terrones}, \citenamefont {Meunier},\ and\ \citenamefont
  {Dresselhaus}}]{Huang14}%
  \BibitemOpen
  \bibfield  {author} {\bibinfo {author} {\bibfnamefont {S.}~\bibnamefont
  {Huang}}, \bibinfo {author} {\bibfnamefont {X.}~\bibnamefont {Ling}},
  \bibinfo {author} {\bibfnamefont {L.}~\bibnamefont {Liang}}, \bibinfo
  {author} {\bibfnamefont {J.}~\bibnamefont {Kong}}, \bibinfo {author}
  {\bibfnamefont {H.}~\bibnamefont {Terrones}}, \bibinfo {author}
  {\bibfnamefont {V.}~\bibnamefont {Meunier}}, \ and\ \bibinfo {author}
  {\bibfnamefont {M.~S.}\ \bibnamefont {Dresselhaus}},\ }\bibfield  {title}
  {\enquote {\bibinfo {title} {Probing the interlayer coupling of twisted
  bilayer {M}o{S}$_2$ using photoluminescence spectroscopy},}\ }\href {\doibase
  10.1021/nl5014597} {\bibfield  {journal} {\bibinfo  {journal} {Nano Letters}\
  }\textbf {\bibinfo {volume} {14}},\ \bibinfo {pages} {5500--5508} (\bibinfo
  {year} {2014})}\BibitemShut {NoStop}%
\bibitem [{\citenamefont {Huang}\ \emph {et~al.}(2016)\citenamefont {Huang},
  \citenamefont {Liang}, \citenamefont {Ling}, \citenamefont {Puretzky},
  \citenamefont {Geohegan}, \citenamefont {Sumpter}, \citenamefont {Kong},
  \citenamefont {Meunier},\ and\ \citenamefont {Dresselhaus}}]{Huang16}%
  \BibitemOpen
  \bibfield  {author} {\bibinfo {author} {\bibfnamefont {S.}~\bibnamefont
  {Huang}}, \bibinfo {author} {\bibfnamefont {L.}~\bibnamefont {Liang}},
  \bibinfo {author} {\bibfnamefont {X.}~\bibnamefont {Ling}}, \bibinfo {author}
  {\bibfnamefont {A.~A.}\ \bibnamefont {Puretzky}}, \bibinfo {author}
  {\bibfnamefont {D.~B.}\ \bibnamefont {Geohegan}}, \bibinfo {author}
  {\bibfnamefont {B.~G.}\ \bibnamefont {Sumpter}}, \bibinfo {author}
  {\bibfnamefont {J.}~\bibnamefont {Kong}}, \bibinfo {author} {\bibfnamefont
  {V.}~\bibnamefont {Meunier}}, \ and\ \bibinfo {author} {\bibfnamefont
  {M.~S.}\ \bibnamefont {Dresselhaus}},\ }\bibfield  {title} {\enquote
  {\bibinfo {title} {Low-frequency interlayer {R}aman modes to probe interface
  of twisted bilayer {M}o{S}$_2$},}\ }\href {\doibase
  10.1021/acs.nanolett.5b05015} {\bibfield  {journal} {\bibinfo  {journal}
  {Nano Letters}\ }\textbf {\bibinfo {volume} {16}},\ \bibinfo {pages}
  {1435--1444} (\bibinfo {year} {2016})}\BibitemShut {NoStop}%
\bibitem [{\citenamefont {Zhang}\ \emph {et~al.}(2017)\citenamefont {Zhang},
  \citenamefont {Chuu}, \citenamefont {Ren}, \citenamefont {Li}, \citenamefont
  {Li}, \citenamefont {Jin}, \citenamefont {Chou},\ and\ \citenamefont
  {Shih}}]{Zhang17}%
  \BibitemOpen
  \bibfield  {author} {\bibinfo {author} {\bibfnamefont {C.}~\bibnamefont
  {Zhang}}, \bibinfo {author} {\bibfnamefont {C.-P.}\ \bibnamefont {Chuu}},
  \bibinfo {author} {\bibfnamefont {X.}~\bibnamefont {Ren}}, \bibinfo {author}
  {\bibfnamefont {M.-Y.}\ \bibnamefont {Li}}, \bibinfo {author} {\bibfnamefont
  {L.-J.}\ \bibnamefont {Li}}, \bibinfo {author} {\bibfnamefont
  {C.}~\bibnamefont {Jin}}, \bibinfo {author} {\bibfnamefont {M.-Y.}\
  \bibnamefont {Chou}}, \ and\ \bibinfo {author} {\bibfnamefont {C.-K.}\
  \bibnamefont {Shih}},\ }\bibfield  {title} {\enquote {\bibinfo {title}
  {Interlayer couplings, {M}oir{\'e} patterns, and {2D} electronic
  superlattices in {M}o{S}$_2$/{WS}e$_2$ hetero-bilayers},}\ }\href {\doibase
  10.1126/sciadv.1601459} {\bibfield  {journal} {\bibinfo  {journal} {Science
  Advances}\ }\textbf {\bibinfo {volume} {3}},\ \bibinfo {pages} {e1601459}
  (\bibinfo {year} {2017})}\BibitemShut {NoStop}%
\bibitem [{\citenamefont {Trainer}\ \emph {et~al.}(2017)\citenamefont
  {Trainer}, \citenamefont {Putilov}, \citenamefont {Di~Giorgio}, \citenamefont
  {Saari}, \citenamefont {Wang}, \citenamefont {Wolak}, \citenamefont
  {Chandrasena}, \citenamefont {Lane}, \citenamefont {Chang}, \citenamefont
  {Jeng}, \citenamefont {Lin}, \citenamefont {Kronast}, \citenamefont {Gray},
  \citenamefont {Xi}, \citenamefont {Nieminen}, \citenamefont {Bansil},\ and\
  \citenamefont {Iavarone}}]{Trainer17}%
  \BibitemOpen
  \bibfield  {author} {\bibinfo {author} {\bibfnamefont {D.~J.}\ \bibnamefont
  {Trainer}}, \bibinfo {author} {\bibfnamefont {A.~V.}\ \bibnamefont
  {Putilov}}, \bibinfo {author} {\bibfnamefont {C.}~\bibnamefont {Di~Giorgio}},
  \bibinfo {author} {\bibfnamefont {T.}~\bibnamefont {Saari}}, \bibinfo
  {author} {\bibfnamefont {B.}~\bibnamefont {Wang}}, \bibinfo {author}
  {\bibfnamefont {M.}~\bibnamefont {Wolak}}, \bibinfo {author} {\bibfnamefont
  {R.~U.}\ \bibnamefont {Chandrasena}}, \bibinfo {author} {\bibfnamefont
  {C.}~\bibnamefont {Lane}}, \bibinfo {author} {\bibfnamefont {T.-R.}\
  \bibnamefont {Chang}}, \bibinfo {author} {\bibfnamefont {H.-T.}\ \bibnamefont
  {Jeng}}, \bibinfo {author} {\bibfnamefont {H.}~\bibnamefont {Lin}}, \bibinfo
  {author} {\bibfnamefont {F.}~\bibnamefont {Kronast}}, \bibinfo {author}
  {\bibfnamefont {A.~X.}\ \bibnamefont {Gray}}, \bibinfo {author}
  {\bibfnamefont {X.}~\bibnamefont {Xi}}, \bibinfo {author} {\bibfnamefont
  {J.}~\bibnamefont {Nieminen}}, \bibinfo {author} {\bibfnamefont
  {A.}~\bibnamefont {Bansil}}, \ and\ \bibinfo {author} {\bibfnamefont
  {M.}~\bibnamefont {Iavarone}},\ }\bibfield  {title} {\enquote {\bibinfo
  {title} {Inter-layer coupling induced valence band edge shift in mono- to
  few-layer {M}o{S}$_2$},}\ }\href {\doibase 10.1038/srep40559} {\bibfield
  {journal} {\bibinfo  {journal} {Scientific Reports}\ }\textbf {\bibinfo
  {volume} {7}},\ \bibinfo {pages} {40559} (\bibinfo {year}
  {2017})}\BibitemShut {NoStop}%
\bibitem [{\citenamefont {Yeh}\ \emph {et~al.}(2016)\citenamefont {Yeh},
  \citenamefont {Jin}, \citenamefont {Zaki}, \citenamefont {Kunstmann},
  \citenamefont {Chenet}, \citenamefont {Arefe}, \citenamefont {Sadowski},
  \citenamefont {Dadap}, \citenamefont {Sutter}, \citenamefont {Hone},\ and\
  \citenamefont {Osgood{,}~Jr.}}]{Yeh16}%
  \BibitemOpen
  \bibfield  {author} {\bibinfo {author} {\bibfnamefont {P.-C.}\ \bibnamefont
  {Yeh}}, \bibinfo {author} {\bibfnamefont {W.}~\bibnamefont {Jin}}, \bibinfo
  {author} {\bibfnamefont {N.}~\bibnamefont {Zaki}}, \bibinfo {author}
  {\bibfnamefont {J.}~\bibnamefont {Kunstmann}}, \bibinfo {author}
  {\bibfnamefont {D.}~\bibnamefont {Chenet}}, \bibinfo {author} {\bibfnamefont
  {G.}~\bibnamefont {Arefe}}, \bibinfo {author} {\bibfnamefont {J.~T.}\
  \bibnamefont {Sadowski}}, \bibinfo {author} {\bibfnamefont {J.~I.}\
  \bibnamefont {Dadap}}, \bibinfo {author} {\bibfnamefont {P.}~\bibnamefont
  {Sutter}}, \bibinfo {author} {\bibfnamefont {J.}~\bibnamefont {Hone}}, \ and\
  \bibinfo {author} {\bibfnamefont {R.~M.}\ \bibnamefont {Osgood{,}~Jr.}},\
  }\bibfield  {title} {\enquote {\bibinfo {title} {Direct measurement of the
  tunable electronic structure of bilayer {M}o{S}$_2$ by interlayer twist},}\
  }\href {\doibase 10.1021/acs.nanolett.5b03883} {\bibfield  {journal}
  {\bibinfo  {journal} {Nano Letters}\ }\textbf {\bibinfo {volume} {16}},\
  \bibinfo {pages} {953--959} (\bibinfo {year} {2016})}\BibitemShut {NoStop}%
\bibitem [{\citenamefont {Lin}\ \emph {et~al.}(2018)\citenamefont {Lin},
  \citenamefont {Tan}, \citenamefont {Wu}, \citenamefont {Chen}, \citenamefont
  {Wang}, \citenamefont {Pan}, \citenamefont {Zhang}, \citenamefont {Cong},
  \citenamefont {Zhang}, \citenamefont {Ji}, \citenamefont {Hu}, \citenamefont
  {Liu},\ and\ \citenamefont {Tan}}]{Lin18}%
  \BibitemOpen
  \bibfield  {author} {\bibinfo {author} {\bibfnamefont {M.-L.}\ \bibnamefont
  {Lin}}, \bibinfo {author} {\bibfnamefont {Q.-H.}\ \bibnamefont {Tan}},
  \bibinfo {author} {\bibfnamefont {J.-B.}\ \bibnamefont {Wu}}, \bibinfo
  {author} {\bibfnamefont {X.-S.}\ \bibnamefont {Chen}}, \bibinfo {author}
  {\bibfnamefont {J.-H.}\ \bibnamefont {Wang}}, \bibinfo {author}
  {\bibfnamefont {Y.-H.}\ \bibnamefont {Pan}}, \bibinfo {author} {\bibfnamefont
  {X.}~\bibnamefont {Zhang}}, \bibinfo {author} {\bibfnamefont
  {X.}~\bibnamefont {Cong}}, \bibinfo {author} {\bibfnamefont {J.}~\bibnamefont
  {Zhang}}, \bibinfo {author} {\bibfnamefont {W.}~\bibnamefont {Ji}}, \bibinfo
  {author} {\bibfnamefont {P.-A.}\ \bibnamefont {Hu}}, \bibinfo {author}
  {\bibfnamefont {K.-H.}\ \bibnamefont {Liu}}, \ and\ \bibinfo {author}
  {\bibfnamefont {P.-H.}\ \bibnamefont {Tan}},\ }\bibfield  {title} {\enquote
  {\bibinfo {title} {Moir{\'e} phonons in twisted bilayer {M}o{S}$_2$},}\
  }\href {\doibase 10.1021/acsnano.8b05006} {\bibfield  {journal} {\bibinfo
  {journal} {ACS Nano}\ }\textbf {\bibinfo {volume} {12}},\ \bibinfo {pages}
  {8770--8780} (\bibinfo {year} {2018})}\BibitemShut {NoStop}%
\bibitem [{\citenamefont {Pan}\ \emph {et~al.}(2018)\citenamefont {Pan},
  \citenamefont {F\"olsch}, \citenamefont {Nie}, \citenamefont {Waters},
  \citenamefont {Lin}, \citenamefont {Jariwala}, \citenamefont {Zhang},
  \citenamefont {Cho}, \citenamefont {Robinson},\ and\ \citenamefont
  {Feenstra}}]{Pan18}%
  \BibitemOpen
  \bibfield  {author} {\bibinfo {author} {\bibfnamefont {Y.}~\bibnamefont
  {Pan}}, \bibinfo {author} {\bibfnamefont {S.}~\bibnamefont {F\"olsch}},
  \bibinfo {author} {\bibfnamefont {Y.}~\bibnamefont {Nie}}, \bibinfo {author}
  {\bibfnamefont {D.}~\bibnamefont {Waters}}, \bibinfo {author} {\bibfnamefont
  {Y.-C.}\ \bibnamefont {Lin}}, \bibinfo {author} {\bibfnamefont
  {B.}~\bibnamefont {Jariwala}}, \bibinfo {author} {\bibfnamefont
  {K.}~\bibnamefont {Zhang}}, \bibinfo {author} {\bibfnamefont
  {K.}~\bibnamefont {Cho}}, \bibinfo {author} {\bibfnamefont {J.~A.}\
  \bibnamefont {Robinson}}, \ and\ \bibinfo {author} {\bibfnamefont {R.~M.}\
  \bibnamefont {Feenstra}},\ }\bibfield  {title} {\enquote {\bibinfo {title}
  {Quantum-confined electronic states arising from the {M}oir\'e pattern of
  {M}o{S}$_2$--{WS}e$_2$ heterobilayers},}\ }\href {\doibase
  10.1021/acs.nanolett.7b05125} {\bibfield  {journal} {\bibinfo  {journal}
  {Nano Letters}\ }\textbf {\bibinfo {volume} {18}},\ \bibinfo {pages}
  {1849--1855} (\bibinfo {year} {2018})}\BibitemShut {NoStop}%
\bibitem [{\citenamefont {Zhang}\ \emph
  {et~al.}(2020{\natexlab{a}})\citenamefont {Zhang}, \citenamefont {Wang},
  \citenamefont {Watanabe}, \citenamefont {Taniguchi}, \citenamefont {Ueno},
  \citenamefont {Tutuc},\ and\ \citenamefont {LeRoy}}]{zhang2019}%
  \BibitemOpen
  \bibfield  {author} {\bibinfo {author} {\bibfnamefont {Z.}~\bibnamefont
  {Zhang}}, \bibinfo {author} {\bibfnamefont {Y.}~\bibnamefont {Wang}},
  \bibinfo {author} {\bibfnamefont {K.}~\bibnamefont {Watanabe}}, \bibinfo
  {author} {\bibfnamefont {T.}~\bibnamefont {Taniguchi}}, \bibinfo {author}
  {\bibfnamefont {K.}~\bibnamefont {Ueno}}, \bibinfo {author} {\bibfnamefont
  {E.}~\bibnamefont {Tutuc}}, \ and\ \bibinfo {author} {\bibfnamefont {B.~J.}\
  \bibnamefont {LeRoy}},\ }\bibfield  {title} {\enquote {\bibinfo {title} {Flat
  bands in twisted bilayer transition metal dichalcogenides},}\ }\href
  {\doibase 10.1038/s41567-020-0958-x} {\bibfield  {journal} {\bibinfo
  {journal} {Nature Physics}\ } (\bibinfo {year} {2020}{\natexlab{a}}),\
  10.1038/s41567-020-0958-x}\BibitemShut {NoStop}%
\bibitem [{\citenamefont {Rold\'an}\ \emph {et~al.}(2014)\citenamefont
  {Rold\'an}, \citenamefont {Silva-Guill\'en}, \citenamefont {L\'opez-Sancho},
  \citenamefont {Guinea}, \citenamefont {Cappelluti},\ and\ \citenamefont
  {Ordej\'on}}]{Roldan14b}%
  \BibitemOpen
  \bibfield  {author} {\bibinfo {author} {\bibfnamefont {R.}~\bibnamefont
  {Rold\'an}}, \bibinfo {author} {\bibfnamefont {J.~A.}\ \bibnamefont
  {Silva-Guill\'en}}, \bibinfo {author} {\bibfnamefont {M.~P.}\ \bibnamefont
  {L\'opez-Sancho}}, \bibinfo {author} {\bibfnamefont {F.}~\bibnamefont
  {Guinea}}, \bibinfo {author} {\bibfnamefont {E.}~\bibnamefont {Cappelluti}},
  \ and\ \bibinfo {author} {\bibfnamefont {P.}~\bibnamefont {Ordej\'on}},\
  }\bibfield  {title} {\enquote {\bibinfo {title} {Electronic properties of
  single-layer and multilayer transition metal dichalcogenides ${MX}_2$ (${M}$
  = {M}o, {W} and ${X}$ = {S}, {S}e)},}\ }\href {\doibase
  10.1002/andp.201400128} {\bibfield  {journal} {\bibinfo  {journal} {Annalen
  der Physik}\ }\textbf {\bibinfo {volume} {526}},\ \bibinfo {pages} {347--357}
  (\bibinfo {year} {2014})}\BibitemShut {NoStop}%
\bibitem [{\citenamefont {Fang}\ \emph {et~al.}(2015)\citenamefont {Fang},
  \citenamefont {Kuate~Defo}, \citenamefont {Shirodkar}, \citenamefont {Lieu},
  \citenamefont {Tritsaris},\ and\ \citenamefont {Kaxiras}}]{Fang15}%
  \BibitemOpen
  \bibfield  {author} {\bibinfo {author} {\bibfnamefont {S.}~\bibnamefont
  {Fang}}, \bibinfo {author} {\bibfnamefont {R.}~\bibnamefont {Kuate~Defo}},
  \bibinfo {author} {\bibfnamefont {S.~N.}\ \bibnamefont {Shirodkar}}, \bibinfo
  {author} {\bibfnamefont {S.}~\bibnamefont {Lieu}}, \bibinfo {author}
  {\bibfnamefont {G.~A.}\ \bibnamefont {Tritsaris}}, \ and\ \bibinfo {author}
  {\bibfnamefont {E.}~\bibnamefont {Kaxiras}},\ }\bibfield  {title} {\enquote
  {\bibinfo {title} {Ab initio tight-binding {H}amiltonian for transition metal
  dichalcogenides},}\ }\href {\doibase 10.1103/PhysRevB.92.205108} {\bibfield
  {journal} {\bibinfo  {journal} {Phys. Rev. B}\ }\textbf {\bibinfo {volume}
  {92}},\ \bibinfo {pages} {205108} (\bibinfo {year} {2015})}\BibitemShut
  {NoStop}%
\bibitem [{\citenamefont {Cao}\ and\ \citenamefont {Li}(2015)}]{Cao15}%
  \BibitemOpen
  \bibfield  {author} {\bibinfo {author} {\bibfnamefont {B.}~\bibnamefont
  {Cao}}\ and\ \bibinfo {author} {\bibfnamefont {T.}~\bibnamefont {Li}},\
  }\bibfield  {title} {\enquote {\bibinfo {title} {Interlayer electronic
  coupling in arbitrarily stacked {M}o{S}$_2$ bilayers controlled by interlayer
  {S}-{S} interaction},}\ }\href {\doibase 10.1021/jp5101736} {\bibfield
  {journal} {\bibinfo  {journal} {The Journal of Physical Chemistry C}\
  }\textbf {\bibinfo {volume} {119}},\ \bibinfo {pages} {1247--1252} (\bibinfo
  {year} {2015})}\BibitemShut {NoStop}%
\bibitem [{\citenamefont {Wang}\ \emph
  {et~al.}(2015{\natexlab{b}})\citenamefont {Wang}, \citenamefont {Chen},\ and\
  \citenamefont {Wang}}]{Wang15}%
  \BibitemOpen
  \bibfield  {author} {\bibinfo {author} {\bibfnamefont {Z.}~\bibnamefont
  {Wang}}, \bibinfo {author} {\bibfnamefont {Q.}~\bibnamefont {Chen}}, \ and\
  \bibinfo {author} {\bibfnamefont {J.}~\bibnamefont {Wang}},\ }\bibfield
  {title} {\enquote {\bibinfo {title} {Electronic structure of twisted bilayers
  of graphene/{M}o{S}$_2$ and {M}o{S}$_2$/{M}o{S}$_2$},}\ }\href {\doibase
  10.1021/jp507751p} {\bibfield  {journal} {\bibinfo  {journal} {The Journal of
  Physical Chemistry C}\ }\textbf {\bibinfo {volume} {119}},\ \bibinfo {pages}
  {4752--4758} (\bibinfo {year} {2015}{\natexlab{b}})}\BibitemShut {NoStop}%
\bibitem [{\citenamefont {Constantinescu}\ and\ \citenamefont
  {Hine}(2015)}]{Constantinescu15}%
  \BibitemOpen
  \bibfield  {author} {\bibinfo {author} {\bibfnamefont {G.~C.}\ \bibnamefont
  {Constantinescu}}\ and\ \bibinfo {author} {\bibfnamefont {N.~D.~M.}\
  \bibnamefont {Hine}},\ }\bibfield  {title} {\enquote {\bibinfo {title}
  {Energy landscape and band-structure tuning in realistic
  {M}o{S}$_{2}$/{M}o{S}e$_{2}$ heterostructures},}\ }\href {\doibase
  10.1103/PhysRevB.91.195416} {\bibfield  {journal} {\bibinfo  {journal} {Phys.
  Rev. B}\ }\textbf {\bibinfo {volume} {91}},\ \bibinfo {pages} {195416}
  (\bibinfo {year} {2015})}\BibitemShut {NoStop}%
\bibitem [{\citenamefont {Tan}\ \emph {et~al.}(2016)\citenamefont {Tan},
  \citenamefont {Chen},\ and\ \citenamefont {Ghosh}}]{Tan16}%
  \BibitemOpen
  \bibfield  {author} {\bibinfo {author} {\bibfnamefont {Y.}~\bibnamefont
  {Tan}}, \bibinfo {author} {\bibfnamefont {F.~W.}\ \bibnamefont {Chen}}, \
  and\ \bibinfo {author} {\bibfnamefont {A.~W.}\ \bibnamefont {Ghosh}},\
  }\bibfield  {title} {\enquote {\bibinfo {title} {First principles study and
  empirical parametrization of twisted bilayer {M}o{S}$_2$ based on
  band-unfolding},}\ }\href {\doibase 10.1063/1.4962438} {\bibfield  {journal}
  {\bibinfo  {journal} {Applied Physics Letters}\ }\textbf {\bibinfo {volume}
  {109}},\ \bibinfo {pages} {101601} (\bibinfo {year} {2016})}\BibitemShut
  {NoStop}%
\bibitem [{\citenamefont {Lu}\ \emph {et~al.}(2017)\citenamefont {Lu},
  \citenamefont {Guo}, \citenamefont {Zhuo}, \citenamefont {Wang},
  \citenamefont {Wu},\ and\ \citenamefont {Zeng}}]{Lu17}%
  \BibitemOpen
  \bibfield  {author} {\bibinfo {author} {\bibfnamefont {N.}~\bibnamefont
  {Lu}}, \bibinfo {author} {\bibfnamefont {H.}~\bibnamefont {Guo}}, \bibinfo
  {author} {\bibfnamefont {Z.}~\bibnamefont {Zhuo}}, \bibinfo {author}
  {\bibfnamefont {L.}~\bibnamefont {Wang}}, \bibinfo {author} {\bibfnamefont
  {X.}~\bibnamefont {Wu}}, \ and\ \bibinfo {author} {\bibfnamefont {X.~C.}\
  \bibnamefont {Zeng}},\ }\bibfield  {title} {\enquote {\bibinfo {title}
  {Twisted {MX}$_2$/{M}o{S}$_2$ heterobilayers: effect of van der {W}aals
  interaction on the electronic structure},}\ }\href {\doibase
  10.1039/C7NR07746G} {\bibfield  {journal} {\bibinfo  {journal} {Nanoscale}\
  }\textbf {\bibinfo {volume} {9}},\ \bibinfo {pages} {19131--19138} (\bibinfo
  {year} {2017})}\BibitemShut {NoStop}%
\bibitem [{\citenamefont {Naik}\ and\ \citenamefont {Jain}(2018)}]{Naik18}%
  \BibitemOpen
  \bibfield  {author} {\bibinfo {author} {\bibfnamefont {M.~H.}\ \bibnamefont
  {Naik}}\ and\ \bibinfo {author} {\bibfnamefont {M.}~\bibnamefont {Jain}},\
  }\bibfield  {title} {\enquote {\bibinfo {title} {Ultraflatbands and shear
  solitons in {M}oir\'e patterns of twisted bilayer transition metal
  dichalcogenides},}\ }\href {\doibase 10.1103/PhysRevLett.121.266401}
  {\bibfield  {journal} {\bibinfo  {journal} {Phys. Rev. Lett.}\ }\textbf
  {\bibinfo {volume} {121}},\ \bibinfo {pages} {266401} (\bibinfo {year}
  {2018})}\BibitemShut {NoStop}%
\bibitem [{\citenamefont {Conte}\ \emph {et~al.}(2019)\citenamefont {Conte},
  \citenamefont {Ninno},\ and\ \citenamefont {Cantele}}]{Conte19}%
  \BibitemOpen
  \bibfield  {author} {\bibinfo {author} {\bibfnamefont {F.}~\bibnamefont
  {Conte}}, \bibinfo {author} {\bibfnamefont {D.}~\bibnamefont {Ninno}}, \ and\
  \bibinfo {author} {\bibfnamefont {G.}~\bibnamefont {Cantele}},\ }\bibfield
  {title} {\enquote {\bibinfo {title} {Electronic properties and interlayer
  coupling of twisted {M}o{S}$_{2}$/{N}b{S}e$_{2}$ heterobilayers},}\ }\href
  {\doibase 10.1103/PhysRevB.99.155429} {\bibfield  {journal} {\bibinfo
  {journal} {Phys. Rev. B}\ }\textbf {\bibinfo {volume} {99}},\ \bibinfo
  {pages} {155429} (\bibinfo {year} {2019})}\BibitemShut {NoStop}%
\bibitem [{\citenamefont {Maity}\ \emph {et~al.}(2019)\citenamefont {Maity},
  \citenamefont {Maiti}, \citenamefont {Krishnamurthy},\ and\ \citenamefont
  {Jain}}]{Maity19}%
  \BibitemOpen
  \bibfield  {author} {\bibinfo {author} {\bibfnamefont {I.}~\bibnamefont
  {Maity}}, \bibinfo {author} {\bibfnamefont {P.~K.}\ \bibnamefont {Maiti}},
  \bibinfo {author} {\bibfnamefont {H.~R.}\ \bibnamefont {Krishnamurthy}}, \
  and\ \bibinfo {author} {\bibfnamefont {M.}~\bibnamefont {Jain}},\ }\href@noop
  {} {\enquote {\bibinfo {title} {Reconstruction of moir\'e lattices in twisted
  transition metal dichalcogenide bilayers},}\ } (\bibinfo {year} {2019}),\
  \Eprint {http://arxiv.org/abs/1912.08702} {arXiv:1912.08702
  [cond-mat.mtrl-sci]} \BibitemShut {NoStop}%
\bibitem [{\citenamefont {Tang}\ \emph {et~al.}(2020)\citenamefont {Tang},
  \citenamefont {Li}, \citenamefont {Li}, \citenamefont {Xu}, \citenamefont
  {Liu}, \citenamefont {Barmak}, \citenamefont {Watanabe}, \citenamefont
  {Taniguchi}, \citenamefont {MacDonald}, \citenamefont {Shan},\ and\
  \citenamefont {Mak}}]{Tang20}%
  \BibitemOpen
  \bibfield  {author} {\bibinfo {author} {\bibfnamefont {Y.}~\bibnamefont
  {Tang}}, \bibinfo {author} {\bibfnamefont {L.}~\bibnamefont {Li}}, \bibinfo
  {author} {\bibfnamefont {T.}~\bibnamefont {Li}}, \bibinfo {author}
  {\bibfnamefont {Y.}~\bibnamefont {Xu}}, \bibinfo {author} {\bibfnamefont
  {S.}~\bibnamefont {Liu}}, \bibinfo {author} {\bibfnamefont {K.}~\bibnamefont
  {Barmak}}, \bibinfo {author} {\bibfnamefont {K.}~\bibnamefont {Watanabe}},
  \bibinfo {author} {\bibfnamefont {T.}~\bibnamefont {Taniguchi}}, \bibinfo
  {author} {\bibfnamefont {A.~H.}\ \bibnamefont {MacDonald}}, \bibinfo {author}
  {\bibfnamefont {J.}~\bibnamefont {Shan}}, \ and\ \bibinfo {author}
  {\bibfnamefont {K.~F.}\ \bibnamefont {Mak}},\ }\bibfield  {title} {\enquote
  {\bibinfo {title} {Simulation of {H}ubbard model physics in
  {WS}e$_2$/{WS}$_2$ moir{\'e} superlattices},}\ }\href {\doibase
  10.1038/s41586-020-2085-3} {\bibfield  {journal} {\bibinfo  {journal}
  {Nature}\ }\textbf {\bibinfo {volume} {579}},\ \bibinfo {pages} {353--358}
  (\bibinfo {year} {2020})}\BibitemShut {NoStop}%
\bibitem [{\citenamefont {Wu}\ \emph {et~al.}(2020)\citenamefont {Wu},
  \citenamefont {Meng}, \citenamefont {Yu},\ and\ \citenamefont {Li}}]{Wu20}%
  \BibitemOpen
  \bibfield  {author} {\bibinfo {author} {\bibfnamefont {J.}~\bibnamefont
  {Wu}}, \bibinfo {author} {\bibfnamefont {L.}~\bibnamefont {Meng}}, \bibinfo
  {author} {\bibfnamefont {J.}~\bibnamefont {Yu}}, \ and\ \bibinfo {author}
  {\bibfnamefont {Y.}~\bibnamefont {Li}},\ }\bibfield  {title} {\enquote
  {\bibinfo {title} {A first-principles study of electronic properties of
  twisted {M}o{T}e$_2$},}\ }\href {\doibase 10.1002/pssb.201900412} {\bibfield
  {journal} {\bibinfo  {journal} {physica status solidi (b)}\ }\textbf
  {\bibinfo {volume} {257}},\ \bibinfo {pages} {1900412} (\bibinfo {year}
  {2020})}\BibitemShut {NoStop}%
\bibitem [{\citenamefont {Lu}\ \emph {et~al.}(2020)\citenamefont {Lu},
  \citenamefont {Carr}, \citenamefont {Larson},\ and\ \citenamefont
  {Kaxiras}}]{Lu20}%
  \BibitemOpen
  \bibfield  {author} {\bibinfo {author} {\bibfnamefont {Z.}~\bibnamefont
  {Lu}}, \bibinfo {author} {\bibfnamefont {S.}~\bibnamefont {Carr}}, \bibinfo
  {author} {\bibfnamefont {D.~T.}\ \bibnamefont {Larson}}, \ and\ \bibinfo
  {author} {\bibfnamefont {E.}~\bibnamefont {Kaxiras}},\ }\href@noop {}
  {\enquote {\bibinfo {title} {Lithium intercalation in {M}o{S}$_2$ bilayers
  and implications for moir\'e flat bands},}\ } (\bibinfo {year} {2020}),\
  \Eprint {http://arxiv.org/abs/2004.00238} {arXiv:2004.00238
  [cond-mat.mes-hall]} \BibitemShut {NoStop}%
\bibitem [{\citenamefont {Xian}\ \emph {et~al.}(2020)\citenamefont {Xian},
  \citenamefont {Claassen}, \citenamefont {Kiese}, \citenamefont {Scherer},
  \citenamefont {Trebst}, \citenamefont {Kennes},\ and\ \citenamefont
  {Rubio}}]{Xian2020}%
  \BibitemOpen
  \bibfield  {author} {\bibinfo {author} {\bibfnamefont {L.}~\bibnamefont
  {Xian}}, \bibinfo {author} {\bibfnamefont {M.}~\bibnamefont {Claassen}},
  \bibinfo {author} {\bibfnamefont {D.}~\bibnamefont {Kiese}}, \bibinfo
  {author} {\bibfnamefont {M.~M.}\ \bibnamefont {Scherer}}, \bibinfo {author}
  {\bibfnamefont {S.}~\bibnamefont {Trebst}}, \bibinfo {author} {\bibfnamefont
  {D.~M.}\ \bibnamefont {Kennes}}, \ and\ \bibinfo {author} {\bibfnamefont
  {A.}~\bibnamefont {Rubio}},\ }\bibfield  {title} {\enquote {\bibinfo {title}
  {Realization of nearly dispersionless bands with strong orbital anisotropy
  from destructive interference in twisted bilayer {M}o{S}$_2$},}\ }\href@noop
  {} {\  (\bibinfo {year} {2020})},\ \Eprint {http://arxiv.org/abs/2004.02964}
  {arXiv:2004.02964 [cond-mat.mes-hall]} \BibitemShut {NoStop}%
\bibitem [{\citenamefont {Pan}\ \emph {et~al.}(2020)\citenamefont {Pan},
  \citenamefont {Wu},\ and\ \citenamefont {Das~Sarma}}]{Pan20}%
  \BibitemOpen
  \bibfield  {author} {\bibinfo {author} {\bibfnamefont {H.}~\bibnamefont
  {Pan}}, \bibinfo {author} {\bibfnamefont {F.}~\bibnamefont {Wu}}, \ and\
  \bibinfo {author} {\bibfnamefont {S.}~\bibnamefont {Das~Sarma}},\ }\bibfield
  {title} {\enquote {\bibinfo {title} {Band topology, hubbard model, heisenberg
  model, and dzyaloshinskii-moriya interaction in twisted bilayer
  ${\mathrm{wse}}_{2}$},}\ }\href {\doibase 10.1103/PhysRevResearch.2.033087}
  {\bibfield  {journal} {\bibinfo  {journal} {Phys. Rev. Research}\ }\textbf
  {\bibinfo {volume} {2}},\ \bibinfo {pages} {033087} (\bibinfo {year}
  {2020})}\BibitemShut {NoStop}%
\bibitem [{\citenamefont {Debbichi}\ \emph {et~al.}(2014)\citenamefont
  {Debbichi}, \citenamefont {Eriksson},\ and\ \citenamefont
  {Leb\`egue}}]{Debbichi14}%
  \BibitemOpen
  \bibfield  {author} {\bibinfo {author} {\bibfnamefont {L.}~\bibnamefont
  {Debbichi}}, \bibinfo {author} {\bibfnamefont {O.}~\bibnamefont {Eriksson}},
  \ and\ \bibinfo {author} {\bibfnamefont {S.}~\bibnamefont {Leb\`egue}},\
  }\bibfield  {title} {\enquote {\bibinfo {title} {Electronic structure of
  two-dimensional transition metal dichalcogenide bilayers from ab initio
  theory},}\ }\href {\doibase 10.1103/PhysRevB.89.205311} {\bibfield  {journal}
  {\bibinfo  {journal} {Phys. Rev. B}\ }\textbf {\bibinfo {volume} {89}},\
  \bibinfo {pages} {205311} (\bibinfo {year} {2014})}\BibitemShut {NoStop}%
\bibitem [{\citenamefont {Peng}\ \emph {et~al.}(2014)\citenamefont {Peng},
  \citenamefont {Huai-Hong}, \citenamefont {Teng},\ and\ \citenamefont
  {Zhi-Dong}}]{Tao14}%
  \BibitemOpen
  \bibfield  {author} {\bibinfo {author} {\bibfnamefont {T.}~\bibnamefont
  {Peng}}, \bibinfo {author} {\bibfnamefont {G.}~\bibnamefont {Huai-Hong}},
  \bibinfo {author} {\bibfnamefont {Y.}~\bibnamefont {Teng}}, \ and\ \bibinfo
  {author} {\bibfnamefont {Z.}~\bibnamefont {Zhi-Dong}},\ }\bibfield  {title}
  {\enquote {\bibinfo {title} {Stacking stability of {M}o{S}$_2$ bilayer: An ab
  initio study},}\ }\href {\doibase 10.1088/1674-1056/23/10/106801} {\bibfield
  {journal} {\bibinfo  {journal} {Chinese Physics B}\ }\textbf {\bibinfo
  {volume} {23}},\ \bibinfo {pages} {106801} (\bibinfo {year}
  {2014})}\BibitemShut {NoStop}%
\bibitem [{\citenamefont {He}\ \emph {et~al.}(2014)\citenamefont {He},
  \citenamefont {Hummer},\ and\ \citenamefont {Franchini}}]{He14}%
  \BibitemOpen
  \bibfield  {author} {\bibinfo {author} {\bibfnamefont {J.}~\bibnamefont
  {He}}, \bibinfo {author} {\bibfnamefont {K.}~\bibnamefont {Hummer}}, \ and\
  \bibinfo {author} {\bibfnamefont {C.}~\bibnamefont {Franchini}},\ }\bibfield
  {title} {\enquote {\bibinfo {title} {Stacking effects on the electronic and
  optical properties of bilayer transition metal dichalcogenides {M}o{S}$_{2}$,
  {M}o{S}e$_{2}$, {WS}$_{2}$, and {WSe}$_{2}$},}\ }\href {\doibase
  10.1103/PhysRevB.89.075409} {\bibfield  {journal} {\bibinfo  {journal} {Phys.
  Rev. B}\ }\textbf {\bibinfo {volume} {89}},\ \bibinfo {pages} {075409}
  (\bibinfo {year} {2014})}\BibitemShut {NoStop}%
\bibitem [{\citenamefont {Sun}\ \emph {et~al.}(2020)\citenamefont {Sun},
  \citenamefont {Luo}, \citenamefont {Li}, \citenamefont {Hong}, \citenamefont
  {Yuan},\ and\ \citenamefont {Zhang}}]{sun2020effects}%
  \BibitemOpen
  \bibfield  {author} {\bibinfo {author} {\bibfnamefont {F.}~\bibnamefont
  {Sun}}, \bibinfo {author} {\bibfnamefont {T.}~\bibnamefont {Luo}}, \bibinfo
  {author} {\bibfnamefont {L.}~\bibnamefont {Li}}, \bibinfo {author}
  {\bibfnamefont {A.}~\bibnamefont {Hong}}, \bibinfo {author} {\bibfnamefont
  {C.}~\bibnamefont {Yuan}}, \ and\ \bibinfo {author} {\bibfnamefont
  {W.}~\bibnamefont {Zhang}},\ }\href@noop {} {\enquote {\bibinfo {title}
  {Effects of magic angle on crystal and electronic structures of bilayer
  transition metal dichalcogenides},}\ } (\bibinfo {year} {2020}),\ \Eprint
  {http://arxiv.org/abs/2003.09872} {arXiv:2003.09872 [cond-mat.mtrl-sci]}
  \BibitemShut {NoStop}%
\bibitem [{\citenamefont {Slater}\ and\ \citenamefont
  {Koster}(1954)}]{Slater54}%
  \BibitemOpen
  \bibfield  {author} {\bibinfo {author} {\bibfnamefont {J.~C.}\ \bibnamefont
  {Slater}}\ and\ \bibinfo {author} {\bibfnamefont {G.~F.}\ \bibnamefont
  {Koster}},\ }\bibfield  {title} {\enquote {\bibinfo {title} {Simplified
  {LCAO} method for the periodic potential problem},}\ }\href {\doibase
  10.1103/PhysRev.94.1498} {\bibfield  {journal} {\bibinfo  {journal} {Phys.
  Rev.}\ }\textbf {\bibinfo {volume} {94}},\ \bibinfo {pages} {1498--1524}
  (\bibinfo {year} {1954})}\BibitemShut {NoStop}%
\bibitem [{\citenamefont {Cappelluti}\ \emph {et~al.}(2013)\citenamefont
  {Cappelluti}, \citenamefont {Rold\'an}, \citenamefont {Silva-Guill\'en},
  \citenamefont {Ordej\'on},\ and\ \citenamefont {Guinea}}]{Cappelluti13}%
  \BibitemOpen
  \bibfield  {author} {\bibinfo {author} {\bibfnamefont {E.}~\bibnamefont
  {Cappelluti}}, \bibinfo {author} {\bibfnamefont {R.}~\bibnamefont
  {Rold\'an}}, \bibinfo {author} {\bibfnamefont {J.~A.}\ \bibnamefont
  {Silva-Guill\'en}}, \bibinfo {author} {\bibfnamefont {P.}~\bibnamefont
  {Ordej\'on}}, \ and\ \bibinfo {author} {\bibfnamefont {F.}~\bibnamefont
  {Guinea}},\ }\bibfield  {title} {\enquote {\bibinfo {title} {Tight-binding
  model and direct-gap/indirect-gap transition in single-layer and multilayer
  {M}o{S}$_{2}$},}\ }\href {\doibase 10.1103/PhysRevB.88.075409} {\bibfield
  {journal} {\bibinfo  {journal} {Phys. Rev. B}\ }\textbf {\bibinfo {volume}
  {88}},\ \bibinfo {pages} {075409} (\bibinfo {year} {2013})}\BibitemShut
  {NoStop}%
\bibitem [{\citenamefont {Rostami}\ \emph {et~al.}(2013)\citenamefont
  {Rostami}, \citenamefont {Moghaddam},\ and\ \citenamefont
  {Asgari}}]{Rostami13}%
  \BibitemOpen
  \bibfield  {author} {\bibinfo {author} {\bibfnamefont {H.}~\bibnamefont
  {Rostami}}, \bibinfo {author} {\bibfnamefont {A.~G.}\ \bibnamefont
  {Moghaddam}}, \ and\ \bibinfo {author} {\bibfnamefont {R.}~\bibnamefont
  {Asgari}},\ }\bibfield  {title} {\enquote {\bibinfo {title} {Effective
  lattice {H}amiltonian for monolayer {M}o{S}$_2$: Tailoring electronic
  structure with perpendicular electric and magnetic fields},}\ }\href
  {\doibase 10.1103/PhysRevB.88.085440} {\bibfield  {journal} {\bibinfo
  {journal} {Phys. Rev. B}\ }\textbf {\bibinfo {volume} {88}},\ \bibinfo
  {pages} {085440} (\bibinfo {year} {2013})}\BibitemShut {NoStop}%
\bibitem [{\citenamefont {Zahid}\ \emph {et~al.}(2013)\citenamefont {Zahid},
  \citenamefont {Liu}, \citenamefont {Zhu}, \citenamefont {Wang},\ and\
  \citenamefont {Guo}}]{Zahid13}%
  \BibitemOpen
  \bibfield  {author} {\bibinfo {author} {\bibfnamefont {F.}~\bibnamefont
  {Zahid}}, \bibinfo {author} {\bibfnamefont {L.}~\bibnamefont {Liu}}, \bibinfo
  {author} {\bibfnamefont {Y.}~\bibnamefont {Zhu}}, \bibinfo {author}
  {\bibfnamefont {J.}~\bibnamefont {Wang}}, \ and\ \bibinfo {author}
  {\bibfnamefont {H.}~\bibnamefont {Guo}},\ }\bibfield  {title} {\enquote
  {\bibinfo {title} {A generic tight-binding model for monolayer, bilayer and
  bulk {M}o{S}$_2$},}\ }\href {\doibase 10.1063/1.4804936} {\bibfield
  {journal} {\bibinfo  {journal} {AIP Advances}\ }\textbf {\bibinfo {volume}
  {3}},\ \bibinfo {pages} {052111} (\bibinfo {year} {2013})}\BibitemShut
  {NoStop}%
\bibitem [{\citenamefont {Ridolfi}\ \emph {et~al.}(2015)\citenamefont
  {Ridolfi}, \citenamefont {Le}, \citenamefont {Rahman}, \citenamefont
  {Mucciolo},\ and\ \citenamefont {Lewenkopf}}]{Ridolfi15}%
  \BibitemOpen
  \bibfield  {author} {\bibinfo {author} {\bibfnamefont {E.}~\bibnamefont
  {Ridolfi}}, \bibinfo {author} {\bibfnamefont {D.}~\bibnamefont {Le}},
  \bibinfo {author} {\bibfnamefont {T.~S.}\ \bibnamefont {Rahman}}, \bibinfo
  {author} {\bibfnamefont {E.~R.}\ \bibnamefont {Mucciolo}}, \ and\ \bibinfo
  {author} {\bibfnamefont {C.~H.}\ \bibnamefont {Lewenkopf}},\ }\bibfield
  {title} {\enquote {\bibinfo {title} {A tight-binding model for {M}o{S}$_2$
  monolayers},}\ }\href {\doibase 10.1088/0953-8984/27/36/365501} {\bibfield
  {journal} {\bibinfo  {journal} {J. Phys.: Condens. Matter}\ }\textbf
  {\bibinfo {volume} {27}},\ \bibinfo {pages} {365501} (\bibinfo {year}
  {2015})}\BibitemShut {NoStop}%
\bibitem [{\citenamefont {Silva-Guill\'en}\ \emph {et~al.}(2016)\citenamefont
  {Silva-Guill\'en}, \citenamefont {San-Jose},\ and\ \citenamefont
  {Rold\'an}}]{SilvaGuillen16}%
  \BibitemOpen
  \bibfield  {author} {\bibinfo {author} {\bibfnamefont {J.~\'A.}\ \bibnamefont
  {Silva-Guill\'en}}, \bibinfo {author} {\bibfnamefont {P.}~\bibnamefont
  {San-Jose}}, \ and\ \bibinfo {author} {\bibfnamefont {R.}~\bibnamefont
  {Rold\'an}},\ }\bibfield  {title} {\enquote {\bibinfo {title} {Electronic
  band structure of transition metal dichalcogenides from ab initio and
  {S}later-{K}oster tight-binding model},}\ }\href {\doibase
  10.3390/app6100284} {\bibfield  {journal} {\bibinfo  {journal} {Applied
  Sciences}\ }\textbf {\bibinfo {volume} {6}},\ \bibinfo {pages} {284}
  (\bibinfo {year} {2016})}\BibitemShut {NoStop}%
\bibitem [{\citenamefont {Trambly~de Laissardi\`ere}\ \emph
  {et~al.}(2012)\citenamefont {Trambly~de Laissardi\`ere}, \citenamefont
  {Mayou},\ and\ \citenamefont {Magaud}}]{Trambly12}%
  \BibitemOpen
  \bibfield  {author} {\bibinfo {author} {\bibfnamefont {G.}~\bibnamefont
  {Trambly~de Laissardi\`ere}}, \bibinfo {author} {\bibfnamefont
  {D.}~\bibnamefont {Mayou}}, \ and\ \bibinfo {author} {\bibfnamefont
  {L.}~\bibnamefont {Magaud}},\ }\bibfield  {title} {\enquote {\bibinfo {title}
  {Numerical studies of confined states in rotated bilayers of graphene},}\
  }\href {\doibase 10.1103/PhysRevB.86.125413} {\bibfield  {journal} {\bibinfo
  {journal} {Phys. Rev. B}\ }\textbf {\bibinfo {volume} {86}},\ \bibinfo
  {pages} {125413} (\bibinfo {year} {2012})}\BibitemShut {NoStop}%
\bibitem [{\citenamefont {Campanera}\ \emph {et~al.}(2007)\citenamefont
  {Campanera}, \citenamefont {Savini}, \citenamefont {Suarez-Martinez},\ and\
  \citenamefont {Heggie}}]{Campanera07}%
  \BibitemOpen
  \bibfield  {author} {\bibinfo {author} {\bibfnamefont {J.~M.}\ \bibnamefont
  {Campanera}}, \bibinfo {author} {\bibfnamefont {G.}~\bibnamefont {Savini}},
  \bibinfo {author} {\bibfnamefont {I.}~\bibnamefont {Suarez-Martinez}}, \ and\
  \bibinfo {author} {\bibfnamefont {M.~I.}\ \bibnamefont {Heggie}},\ }\bibfield
   {title} {\enquote {\bibinfo {title} {Density functional calculations on the
  intricacies of {M}oir\'e patterns on graphite},}\ }\href {\doibase
  10.1103/PhysRevB.75.235449} {\bibfield  {journal} {\bibinfo  {journal} {Phys.
  Rev. B}\ }\textbf {\bibinfo {volume} {75}},\ \bibinfo {pages} {235449}
  (\bibinfo {year} {2007})}\BibitemShut {NoStop}%
\bibitem [{\citenamefont {Mele}(2010)}]{Mele10}%
  \BibitemOpen
  \bibfield  {author} {\bibinfo {author} {\bibfnamefont {E.~J.}\ \bibnamefont
  {Mele}},\ }\bibfield  {title} {\enquote {\bibinfo {title} {Commensuration and
  interlayer coherence in twisted bilayer graphene},}\ }\href {\doibase
  10.1103/PhysRevB.81.161405} {\bibfield  {journal} {\bibinfo  {journal} {Phys.
  Rev. B}\ }\textbf {\bibinfo {volume} {81}},\ \bibinfo {pages} {161405(R)}
  (\bibinfo {year} {2010})}\BibitemShut {NoStop}%
\bibitem [{\citenamefont {Huisman}\ \emph {et~al.}(1971)\citenamefont
  {Huisman}, \citenamefont {de~Jonge}, \citenamefont {Haas},\ and\
  \citenamefont {Jellinek}}]{Huisman71}%
  \BibitemOpen
  \bibfield  {author} {\bibinfo {author} {\bibfnamefont {R.}~\bibnamefont
  {Huisman}}, \bibinfo {author} {\bibfnamefont {R.}~\bibnamefont {de~Jonge}},
  \bibinfo {author} {\bibfnamefont {C.}~\bibnamefont {Haas}}, \ and\ \bibinfo
  {author} {\bibfnamefont {F.}~\bibnamefont {Jellinek}},\ }\bibfield  {title}
  {\enquote {\bibinfo {title} {Trigonal-prismatic coordination in solid
  compounds of transition metals},}\ }\href {\doibase
  10.1016/0022-4596(71)90007-7} {\bibfield  {journal} {\bibinfo  {journal}
  {Journal of Solid State Chemistry}\ }\textbf {\bibinfo {volume} {3}},\
  \bibinfo {pages} {56--66} (\bibinfo {year} {1971})}\BibitemShut {NoStop}%
\bibitem [{sup()}]{supplementaryMat}%
  \BibitemOpen
  \href@noop {} {}\bibinfo {note} {See Supplemental Material (page 8) the
  commensurate moir{\'e} structures that have been used in the present work,
  details and complementary results on our DFT calculations, Slater-Koster
  parameters, and complementary tight-binding results.}\BibitemShut {Stop}%
\bibitem [{\citenamefont {Nam}\ and\ \citenamefont {Koshino}(2017)}]{Nam17}%
  \BibitemOpen
  \bibfield  {author} {\bibinfo {author} {\bibfnamefont {N.~N.~T.}\
  \bibnamefont {Nam}}\ and\ \bibinfo {author} {\bibfnamefont {M.}~\bibnamefont
  {Koshino}},\ }\bibfield  {title} {\enquote {\bibinfo {title} {Lattice
  relaxation and energy band modulation in twisted bilayer graphene},}\ }\href
  {\doibase 10.1103/PhysRevB.96.075311} {\bibfield  {journal} {\bibinfo
  {journal} {Phys. Rev. B}\ }\textbf {\bibinfo {volume} {96}},\ \bibinfo
  {pages} {075311} (\bibinfo {year} {2017})}\BibitemShut {NoStop}%
\bibitem [{\citenamefont {Hohenberg}\ and\ \citenamefont
  {Kohn}(1964)}]{Hohenberg64}%
  \BibitemOpen
  \bibfield  {author} {\bibinfo {author} {\bibfnamefont {P.}~\bibnamefont
  {Hohenberg}}\ and\ \bibinfo {author} {\bibfnamefont {W.}~\bibnamefont
  {Kohn}},\ }\bibfield  {title} {\enquote {\bibinfo {title} {Inhomogeneous
  electron gas},}\ }\href {\doibase 10.1103/PhysRev.136.B864} {\bibfield
  {journal} {\bibinfo  {journal} {Phys. Rev.}\ }\textbf {\bibinfo {volume}
  {136}},\ \bibinfo {pages} {B864--B871} (\bibinfo {year} {1964})}\BibitemShut
  {NoStop}%
\bibitem [{\citenamefont {Kohn}\ and\ \citenamefont {Sham}(1965)}]{Kohn65}%
  \BibitemOpen
  \bibfield  {author} {\bibinfo {author} {\bibfnamefont {W.}~\bibnamefont
  {Kohn}}\ and\ \bibinfo {author} {\bibfnamefont {L.~J.}\ \bibnamefont
  {Sham}},\ }\bibfield  {title} {\enquote {\bibinfo {title} {Self-consistent
  equations including exchange and correlation effects},}\ }\href {\doibase
  10.1103/PhysRev.140.A1133} {\bibfield  {journal} {\bibinfo  {journal} {Phys.
  Rev.}\ }\textbf {\bibinfo {volume} {140}},\ \bibinfo {pages} {A1133--A1138}
  (\bibinfo {year} {1965})}\BibitemShut {NoStop}%
\bibitem [{\citenamefont {Gonze}\ \emph {et~al.}(2002)\citenamefont {Gonze},
  \citenamefont {Beuken}, \citenamefont {Caracas}, \citenamefont {Detraux},
  \citenamefont {Fuchs}, \citenamefont {Rignanese}, \citenamefont {Sindic},
  \citenamefont {Verstraete}, \citenamefont {Zerah}, \citenamefont {Jollet},
  \citenamefont {Torrent}, \citenamefont {Roy}, \citenamefont {Mikami},
  \citenamefont {Ghosez}, \citenamefont {Raty},\ and\ \citenamefont
  {Allan}}]{Gonze02}%
  \BibitemOpen
  \bibfield  {author} {\bibinfo {author} {\bibfnamefont {X.}~\bibnamefont
  {Gonze}}, \bibinfo {author} {\bibfnamefont {J.-M.}\ \bibnamefont {Beuken}},
  \bibinfo {author} {\bibfnamefont {R.}~\bibnamefont {Caracas}}, \bibinfo
  {author} {\bibfnamefont {F.}~\bibnamefont {Detraux}}, \bibinfo {author}
  {\bibfnamefont {M.}~\bibnamefont {Fuchs}}, \bibinfo {author} {\bibfnamefont
  {G.-M.}\ \bibnamefont {Rignanese}}, \bibinfo {author} {\bibfnamefont
  {L.}~\bibnamefont {Sindic}}, \bibinfo {author} {\bibfnamefont
  {M.}~\bibnamefont {Verstraete}}, \bibinfo {author} {\bibfnamefont
  {G.}~\bibnamefont {Zerah}}, \bibinfo {author} {\bibfnamefont
  {F.}~\bibnamefont {Jollet}}, \bibinfo {author} {\bibfnamefont
  {M.}~\bibnamefont {Torrent}}, \bibinfo {author} {\bibfnamefont
  {A.}~\bibnamefont {Roy}}, \bibinfo {author} {\bibfnamefont {M.}~\bibnamefont
  {Mikami}}, \bibinfo {author} {\bibfnamefont {Ph.}\ \bibnamefont {Ghosez}},
  \bibinfo {author} {\bibfnamefont {J.-Y.}\ \bibnamefont {Raty}}, \ and\
  \bibinfo {author} {\bibfnamefont {D.C.}\ \bibnamefont {Allan}},\ }\bibfield
  {title} {\enquote {\bibinfo {title} {First-principles computation of material
  properties: the {ABINIT} software project},}\ }\href {\doibase
  https://doi.org/10.1016/S0927-0256(02)00325-7} {\bibfield  {journal}
  {\bibinfo  {journal} {Computational Materials Science}\ }\textbf {\bibinfo
  {volume} {25}},\ \bibinfo {pages} {478--492} (\bibinfo {year}
  {2002})}\BibitemShut {NoStop}%
\bibitem [{\citenamefont {Gonze}\ \emph {et~al.}(2009)\citenamefont {Gonze},
  \citenamefont {Amadon}, \citenamefont {Anglade}, \citenamefont {Beuken},
  \citenamefont {Bottin}, \citenamefont {Boulanger}, \citenamefont {Bruneval},
  \citenamefont {Caliste}, \citenamefont {Caracas}, \citenamefont {C\^ot\'e},
  \citenamefont {Deutsch}, \citenamefont {Genovese}, \citenamefont {Ghosez},
  \citenamefont {Giantomassi}, \citenamefont {Goedecker}, \citenamefont
  {Hamann}, \citenamefont {Hermet}, \citenamefont {Jollet}, \citenamefont
  {Jomard}, \citenamefont {Leroux}, \citenamefont {Mancini}, \citenamefont
  {Mazevet}, \citenamefont {Oliveira}, \citenamefont {Onida}, \citenamefont
  {Pouillon}, \citenamefont {Rangel}, \citenamefont {Rignanese}, \citenamefont
  {Sangalli}, \citenamefont {Shaltaf}, \citenamefont {Torrent}, \citenamefont
  {Verstraete}, \citenamefont {Zerah},\ and\ \citenamefont
  {Zwanziger}}]{Gonze09}%
  \BibitemOpen
  \bibfield  {author} {\bibinfo {author} {\bibfnamefont {X.}~\bibnamefont
  {Gonze}}, \bibinfo {author} {\bibfnamefont {B.}~\bibnamefont {Amadon}},
  \bibinfo {author} {\bibfnamefont {P.-M.}\ \bibnamefont {Anglade}}, \bibinfo
  {author} {\bibfnamefont {J.-M.}\ \bibnamefont {Beuken}}, \bibinfo {author}
  {\bibfnamefont {F.}~\bibnamefont {Bottin}}, \bibinfo {author} {\bibfnamefont
  {P.}~\bibnamefont {Boulanger}}, \bibinfo {author} {\bibfnamefont
  {F.}~\bibnamefont {Bruneval}}, \bibinfo {author} {\bibfnamefont
  {D.}~\bibnamefont {Caliste}}, \bibinfo {author} {\bibfnamefont
  {R.}~\bibnamefont {Caracas}}, \bibinfo {author} {\bibfnamefont
  {M.}~\bibnamefont {C\^ot\'e}}, \bibinfo {author} {\bibfnamefont
  {T.}~\bibnamefont {Deutsch}}, \bibinfo {author} {\bibfnamefont
  {L.}~\bibnamefont {Genovese}}, \bibinfo {author} {\bibfnamefont {Ph.}\
  \bibnamefont {Ghosez}}, \bibinfo {author} {\bibfnamefont {M.}~\bibnamefont
  {Giantomassi}}, \bibinfo {author} {\bibfnamefont {S.}~\bibnamefont
  {Goedecker}}, \bibinfo {author} {\bibfnamefont {D.~R.}\ \bibnamefont
  {Hamann}}, \bibinfo {author} {\bibfnamefont {P.}~\bibnamefont {Hermet}},
  \bibinfo {author} {\bibfnamefont {F.}~\bibnamefont {Jollet}}, \bibinfo
  {author} {\bibfnamefont {G.}~\bibnamefont {Jomard}}, \bibinfo {author}
  {\bibfnamefont {S.}~\bibnamefont {Leroux}}, \bibinfo {author} {\bibfnamefont
  {M.}~\bibnamefont {Mancini}}, \bibinfo {author} {\bibfnamefont
  {S.}~\bibnamefont {Mazevet}}, \bibinfo {author} {\bibfnamefont {M.~J.~T.}\
  \bibnamefont {Oliveira}}, \bibinfo {author} {\bibfnamefont {G.}~\bibnamefont
  {Onida}}, \bibinfo {author} {\bibfnamefont {Y.}~\bibnamefont {Pouillon}},
  \bibinfo {author} {\bibfnamefont {T.}~\bibnamefont {Rangel}}, \bibinfo
  {author} {\bibfnamefont {G.-M.}\ \bibnamefont {Rignanese}}, \bibinfo {author}
  {\bibfnamefont {D.}~\bibnamefont {Sangalli}}, \bibinfo {author}
  {\bibfnamefont {R.}~\bibnamefont {Shaltaf}}, \bibinfo {author} {\bibfnamefont
  {M.}~\bibnamefont {Torrent}}, \bibinfo {author} {\bibfnamefont {M.~J.}\
  \bibnamefont {Verstraete}}, \bibinfo {author} {\bibfnamefont
  {G.}~\bibnamefont {Zerah}}, \ and\ \bibinfo {author} {\bibfnamefont {J.~W.}\
  \bibnamefont {Zwanziger}},\ }\bibfield  {title} {\enquote {\bibinfo {title}
  {{ABINIT:} first-principles approach to material and nanosystem
  properties},}\ }\href {\doibase https://doi.org/10.1016/j.cpc.2009.07.007}
  {\bibfield  {journal} {\bibinfo  {journal} {Comp. Phys. Commun.}\ }\textbf
  {\bibinfo {volume} {180}},\ \bibinfo {pages} {2582--2615} (\bibinfo {year}
  {2009})}\BibitemShut {NoStop}%
\bibitem [{\citenamefont {Gonze}\ \emph {et~al.}(2016)\citenamefont {Gonze},
  \citenamefont {Jollet}, \citenamefont {Araujo}, \citenamefont {Adams},
  \citenamefont {Amadon}, \citenamefont {Applencourt}, \citenamefont {Audouze},
  \citenamefont {Beuken}, \citenamefont {Bieder}, \citenamefont {Bokhanchuk},
  \citenamefont {Bousquet}, \citenamefont {Bruneval}, \citenamefont {Caliste},
  \citenamefont {C\^ot\'e}, \citenamefont {Dahm}, \citenamefont {Pieve},
  \citenamefont {Delaveau}, \citenamefont {Gennaro}, \citenamefont {Dorado},
  \citenamefont {Espejo}, \citenamefont {Geneste}, \citenamefont {Genovese},
  \citenamefont {Gerossier}, \citenamefont {Giantomassi}, \citenamefont
  {Gillet}, \citenamefont {Hamann}, \citenamefont {He}, \citenamefont {Jomard},
  \citenamefont {Janssen}, \citenamefont {Roux}, \citenamefont {Levitt},
  \citenamefont {Lherbier}, \citenamefont {Liu}, \citenamefont
  {Luka\v{c}evi\'{c}}, \citenamefont {Martin}, \citenamefont {Martins},
  \citenamefont {Oliveira}, \citenamefont {Ponc\'e}, \citenamefont {Pouillon},
  \citenamefont {Rangel}, \citenamefont {Rignanese}, \citenamefont {Romero},
  \citenamefont {Rousseau}, \citenamefont {Rubel}, \citenamefont {Shukri},
  \citenamefont {Stankovski}, \citenamefont {Torrent}, \citenamefont {Setten},
  \citenamefont {Troeye}, \citenamefont {Verstraete}, \citenamefont
  {Waroquiers}, \citenamefont {Wiktor}, \citenamefont {Xu}, \citenamefont
  {Zhou},\ and\ \citenamefont {Zwanziger}}]{Gonze16}%
  \BibitemOpen
  \bibfield  {author} {\bibinfo {author} {\bibfnamefont {X.}~\bibnamefont
  {Gonze}}, \bibinfo {author} {\bibfnamefont {F.}~\bibnamefont {Jollet}},
  \bibinfo {author} {\bibfnamefont {F.~Abreu}\ \bibnamefont {Araujo}}, \bibinfo
  {author} {\bibfnamefont {D.}~\bibnamefont {Adams}}, \bibinfo {author}
  {\bibfnamefont {B.}~\bibnamefont {Amadon}}, \bibinfo {author} {\bibfnamefont
  {T.}~\bibnamefont {Applencourt}}, \bibinfo {author} {\bibfnamefont
  {C.}~\bibnamefont {Audouze}}, \bibinfo {author} {\bibfnamefont {J.-M.}\
  \bibnamefont {Beuken}}, \bibinfo {author} {\bibfnamefont {J.}~\bibnamefont
  {Bieder}}, \bibinfo {author} {\bibfnamefont {A.}~\bibnamefont {Bokhanchuk}},
  \bibinfo {author} {\bibfnamefont {E.}~\bibnamefont {Bousquet}}, \bibinfo
  {author} {\bibfnamefont {F.}~\bibnamefont {Bruneval}}, \bibinfo {author}
  {\bibfnamefont {D.}~\bibnamefont {Caliste}}, \bibinfo {author} {\bibfnamefont
  {M.}~\bibnamefont {C\^ot\'e}}, \bibinfo {author} {\bibfnamefont
  {F.}~\bibnamefont {Dahm}}, \bibinfo {author} {\bibfnamefont {F.~Da}\
  \bibnamefont {Pieve}}, \bibinfo {author} {\bibfnamefont {M.}~\bibnamefont
  {Delaveau}}, \bibinfo {author} {\bibfnamefont {M.~Di}\ \bibnamefont
  {Gennaro}}, \bibinfo {author} {\bibfnamefont {B.}~\bibnamefont {Dorado}},
  \bibinfo {author} {\bibfnamefont {C.}~\bibnamefont {Espejo}}, \bibinfo
  {author} {\bibfnamefont {G.}~\bibnamefont {Geneste}}, \bibinfo {author}
  {\bibfnamefont {L.}~\bibnamefont {Genovese}}, \bibinfo {author}
  {\bibfnamefont {A.}~\bibnamefont {Gerossier}}, \bibinfo {author}
  {\bibfnamefont {M.}~\bibnamefont {Giantomassi}}, \bibinfo {author}
  {\bibfnamefont {Y.}~\bibnamefont {Gillet}}, \bibinfo {author} {\bibfnamefont
  {D.~R.}\ \bibnamefont {Hamann}}, \bibinfo {author} {\bibfnamefont
  {L.}~\bibnamefont {He}}, \bibinfo {author} {\bibfnamefont {G.}~\bibnamefont
  {Jomard}}, \bibinfo {author} {\bibfnamefont {J.~Laflamme}\ \bibnamefont
  {Janssen}}, \bibinfo {author} {\bibfnamefont {S.~Le}\ \bibnamefont {Roux}},
  \bibinfo {author} {\bibfnamefont {A.}~\bibnamefont {Levitt}}, \bibinfo
  {author} {\bibfnamefont {A.}~\bibnamefont {Lherbier}}, \bibinfo {author}
  {\bibfnamefont {F.}~\bibnamefont {Liu}}, \bibinfo {author} {\bibfnamefont
  {I.}~\bibnamefont {Luka\v{c}evi\'{c}}}, \bibinfo {author} {\bibfnamefont
  {A.}~\bibnamefont {Martin}}, \bibinfo {author} {\bibfnamefont
  {C.}~\bibnamefont {Martins}}, \bibinfo {author} {\bibfnamefont {M.~J.~T.}\
  \bibnamefont {Oliveira}}, \bibinfo {author} {\bibfnamefont {S.}~\bibnamefont
  {Ponc\'e}}, \bibinfo {author} {\bibfnamefont {Y.}~\bibnamefont {Pouillon}},
  \bibinfo {author} {\bibfnamefont {T.}~\bibnamefont {Rangel}}, \bibinfo
  {author} {\bibfnamefont {G.-M.}\ \bibnamefont {Rignanese}}, \bibinfo {author}
  {\bibfnamefont {A.~H.}\ \bibnamefont {Romero}}, \bibinfo {author}
  {\bibfnamefont {B.}~\bibnamefont {Rousseau}}, \bibinfo {author}
  {\bibfnamefont {O.}~\bibnamefont {Rubel}}, \bibinfo {author} {\bibfnamefont
  {A.~A.}\ \bibnamefont {Shukri}}, \bibinfo {author} {\bibfnamefont
  {M.}~\bibnamefont {Stankovski}}, \bibinfo {author} {\bibfnamefont
  {M.}~\bibnamefont {Torrent}}, \bibinfo {author} {\bibfnamefont {M.~J.~Van}\
  \bibnamefont {Setten}}, \bibinfo {author} {\bibfnamefont {B.~Van}\
  \bibnamefont {Troeye}}, \bibinfo {author} {\bibfnamefont {M.~J.}\
  \bibnamefont {Verstraete}}, \bibinfo {author} {\bibfnamefont
  {D.}~\bibnamefont {Waroquiers}}, \bibinfo {author} {\bibfnamefont
  {J.}~\bibnamefont {Wiktor}}, \bibinfo {author} {\bibfnamefont
  {B.}~\bibnamefont {Xu}}, \bibinfo {author} {\bibfnamefont {A.}~\bibnamefont
  {Zhou}}, \ and\ \bibinfo {author} {\bibfnamefont {J.~W.}\ \bibnamefont
  {Zwanziger}},\ }\bibfield  {title} {\enquote {\bibinfo {title} {Recent
  developments in the {ABINIT} software package},}\ }\href {\doibase
  https://doi.org/10.1016/j.cpc.2016.04.003} {\bibfield  {journal} {\bibinfo
  {journal} {Comp. Phys. Commun.}\ }\textbf {\bibinfo {volume} {205}},\
  \bibinfo {pages} {106--131} (\bibinfo {year} {2016})}\BibitemShut {NoStop}%
\bibitem [{\citenamefont {Zupan}\ \emph {et~al.}(1998)\citenamefont {Zupan},
  \citenamefont {Blaha}, \citenamefont {Schwarz},\ and\ \citenamefont
  {Perdew}}]{Zupan98}%
  \BibitemOpen
  \bibfield  {author} {\bibinfo {author} {\bibfnamefont {A.}~\bibnamefont
  {Zupan}}, \bibinfo {author} {\bibfnamefont {P.}~\bibnamefont {Blaha}},
  \bibinfo {author} {\bibfnamefont {K.}~\bibnamefont {Schwarz}}, \ and\
  \bibinfo {author} {\bibfnamefont {J.~P.}\ \bibnamefont {Perdew}},\ }\bibfield
   {title} {\enquote {\bibinfo {title} {Pressure-induced phase transitions in
  solid {S}i, {S}i{O}$_{2}$, and {F}e: Performance of local-spin-density and
  generalized-gradient-approximation density functionals},}\ }\href {\doibase
  10.1103/PhysRevB.58.11266} {\bibfield  {journal} {\bibinfo  {journal} {Phys.
  Rev. B}\ }\textbf {\bibinfo {volume} {58}},\ \bibinfo {pages} {11266--11272}
  (\bibinfo {year} {1998})}\BibitemShut {NoStop}%
\bibitem [{\citenamefont {Jones}\ and\ \citenamefont
  {Gunnarsson}(1989)}]{Jones89}%
  \BibitemOpen
  \bibfield  {author} {\bibinfo {author} {\bibfnamefont {R.~O.}\ \bibnamefont
  {Jones}}\ and\ \bibinfo {author} {\bibfnamefont {O.}~\bibnamefont
  {Gunnarsson}},\ }\bibfield  {title} {\enquote {\bibinfo {title} {The density
  functional formalism, its applications and prospects},}\ }\href {\doibase
  10.1103/RevModPhys.61.689} {\bibfield  {journal} {\bibinfo  {journal} {Rev.
  Mod. Phys.}\ }\textbf {\bibinfo {volume} {61}},\ \bibinfo {pages} {689--746}
  (\bibinfo {year} {1989})}\BibitemShut {NoStop}%
\bibitem [{\citenamefont {Perdew}\ \emph {et~al.}(1996)\citenamefont {Perdew},
  \citenamefont {Burke},\ and\ \citenamefont {Ernzerhof}}]{Perdew96}%
  \BibitemOpen
  \bibfield  {author} {\bibinfo {author} {\bibfnamefont {J.~P.}\ \bibnamefont
  {Perdew}}, \bibinfo {author} {\bibfnamefont {K.}~\bibnamefont {Burke}}, \
  and\ \bibinfo {author} {\bibfnamefont {M.}~\bibnamefont {Ernzerhof}},\
  }\bibfield  {title} {\enquote {\bibinfo {title} {Generalized gradient
  approximation made simple},}\ }\href {\doibase 10.1103/PhysRevLett.77.3865}
  {\bibfield  {journal} {\bibinfo  {journal} {Phys. Rev. Lett.}\ }\textbf
  {\bibinfo {volume} {77}},\ \bibinfo {pages} {3865--3868} (\bibinfo {year}
  {1996})}\BibitemShut {NoStop}%
\bibitem [{\citenamefont {Mehl}\ and\ \citenamefont
  {Papaconstantopoulos}(1996)}]{Mehl96}%
  \BibitemOpen
  \bibfield  {author} {\bibinfo {author} {\bibfnamefont {M.~J.}\ \bibnamefont
  {Mehl}}\ and\ \bibinfo {author} {\bibfnamefont {D.~A.}\ \bibnamefont
  {Papaconstantopoulos}},\ }\bibfield  {title} {\enquote {\bibinfo {title}
  {Applications of a tight-binding total-energy method for transition and noble
  metals: Elastic constants, vacancies, and surfaces of monatomic metals},}\
  }\href {\doibase 10.1103/PhysRevB.54.4519} {\bibfield  {journal} {\bibinfo
  {journal} {Phys. Rev. B}\ }\textbf {\bibinfo {volume} {54}},\ \bibinfo
  {pages} {4519--4530} (\bibinfo {year} {1996})}\BibitemShut {NoStop}%
\bibitem [{\citenamefont {Brihuega}\ \emph {et~al.}(2012)\citenamefont
  {Brihuega}, \citenamefont {Mallet}, \citenamefont {Gonz\'alez-Herrero},
  \citenamefont {Trambly~de Laissardi\`ere}, \citenamefont {Ugeda},
  \citenamefont {Magaud}, \citenamefont {G\'omez-Rodr\'{\i}guez}, \citenamefont
  {Yndur\'ain},\ and\ \citenamefont {Veuillen}}]{Brihuega12}%
  \BibitemOpen
  \bibfield  {author} {\bibinfo {author} {\bibfnamefont {I.}~\bibnamefont
  {Brihuega}}, \bibinfo {author} {\bibfnamefont {P.}~\bibnamefont {Mallet}},
  \bibinfo {author} {\bibfnamefont {H.}~\bibnamefont {Gonz\'alez-Herrero}},
  \bibinfo {author} {\bibfnamefont {G.}~\bibnamefont {Trambly~de
  Laissardi\`ere}}, \bibinfo {author} {\bibfnamefont {M.~M.}\ \bibnamefont
  {Ugeda}}, \bibinfo {author} {\bibfnamefont {L.}~\bibnamefont {Magaud}},
  \bibinfo {author} {\bibfnamefont {J.~M.}\ \bibnamefont
  {G\'omez-Rodr\'{\i}guez}}, \bibinfo {author} {\bibfnamefont {F.}~\bibnamefont
  {Yndur\'ain}}, \ and\ \bibinfo {author} {\bibfnamefont {J.-Y.}\ \bibnamefont
  {Veuillen}},\ }\bibfield  {title} {\enquote {\bibinfo {title} {Unraveling the
  intrinsic and robust nature of van {H}ove singularities in twisted bilayer
  graphene by scanning tunneling microscopy and theoretical analysis},}\ }\href
  {\doibase 10.1103/PhysRevLett.109.196802} {\bibfield  {journal} {\bibinfo
  {journal} {Phys. Rev. Lett.}\ }\textbf {\bibinfo {volume} {109}},\ \bibinfo
  {pages} {196802} (\bibinfo {year} {2012})}\BibitemShut {NoStop}%
\bibitem [{\citenamefont {Lopes~dos Santos}\ \emph {et~al.}(2012)\citenamefont
  {Lopes~dos Santos}, \citenamefont {Peres},\ and\ \citenamefont
  {Castro~Neto}}]{LopesdosSantos12}%
  \BibitemOpen
  \bibfield  {author} {\bibinfo {author} {\bibfnamefont {J.~M.~B.}\
  \bibnamefont {Lopes~dos Santos}}, \bibinfo {author} {\bibfnamefont
  {N.~M.~R.}\ \bibnamefont {Peres}}, \ and\ \bibinfo {author} {\bibfnamefont
  {A.~H.}\ \bibnamefont {Castro~Neto}},\ }\bibfield  {title} {\enquote
  {\bibinfo {title} {Continuum model of the twisted graphene bilayer},}\ }\href
  {\doibase 10.1103/PhysRevB.86.155449} {\bibfield  {journal} {\bibinfo
  {journal} {Phys. Rev. B}\ }\textbf {\bibinfo {volume} {86}},\ \bibinfo
  {pages} {155449} (\bibinfo {year} {2012})}\BibitemShut {NoStop}%
\bibitem [{\citenamefont {Namarvar}\ \emph {et~al.}(2020)\citenamefont
  {Namarvar}, \citenamefont {Missaoui}, \citenamefont {Magaud}, \citenamefont
  {Mayou},\ and\ \citenamefont {Trambly~de Laissardi\`ere}}]{Namarvar20}%
  \BibitemOpen
  \bibfield  {author} {\bibinfo {author} {\bibfnamefont {O.~F.}\ \bibnamefont
  {Namarvar}}, \bibinfo {author} {\bibfnamefont {A.}~\bibnamefont {Missaoui}},
  \bibinfo {author} {\bibfnamefont {L.}~\bibnamefont {Magaud}}, \bibinfo
  {author} {\bibfnamefont {D.}~\bibnamefont {Mayou}}, \ and\ \bibinfo {author}
  {\bibfnamefont {G.}~\bibnamefont {Trambly~de Laissardi\`ere}},\ }\bibfield
  {title} {\enquote {\bibinfo {title} {Electronic structure and quantum
  transport in twisted bilayer graphene with resonant scatterers},}\ }\href
  {\doibase 10.1103/PhysRevB.101.245407} {\bibfield  {journal} {\bibinfo
  {journal} {Phys. Rev. B}\ }\textbf {\bibinfo {volume} {101}},\ \bibinfo
  {pages} {245407} (\bibinfo {year} {2020})}\BibitemShut {NoStop}%
\bibitem [{\citenamefont {Zhan}\ \emph {et~al.}(2020)\citenamefont {Zhan},
  \citenamefont {Zhang}, \citenamefont {Yu}, \citenamefont {Guinea},
  \citenamefont {Silva-Guill\'en},\ and\ \citenamefont {Yuan}}]{Zhang2020a}%
  \BibitemOpen
  \bibfield  {author} {\bibinfo {author} {\bibfnamefont {Z.}~\bibnamefont
  {Zhan}}, \bibinfo {author} {\bibfnamefont {Y.}~\bibnamefont {Zhang}},
  \bibinfo {author} {\bibfnamefont {G.}~\bibnamefont {Yu}}, \bibinfo {author}
  {\bibfnamefont {F}~\bibnamefont {Guinea}}, \bibinfo {author} {\bibfnamefont
  {J.~\'A.}\ \bibnamefont {Silva-Guill\'en}}, \ and\ \bibinfo {author}
  {\bibfnamefont {S.}~\bibnamefont {Yuan}},\ }\bibfield  {title} {\enquote
  {\bibinfo {title} {Multi-ultraflatbands tunability and effect of spin-orbit
  coupling in twisted bilayer transition metal dichalcogenides},}\ }\href@noop
  {} {\  (\bibinfo {year} {2020})},\ \Eprint {http://arxiv.org/abs/2005.13868}
  {arXiv:2005.13868 [cond-mat.mes-hall]} \BibitemShut {NoStop}%
\bibitem [{\citenamefont {Zhang}\ \emph
  {et~al.}(2020{\natexlab{b}})\citenamefont {Zhang}, \citenamefont {Zhan},
  \citenamefont {Guinea}, \citenamefont {Silva-Guill\'en},\ and\ \citenamefont
  {Yuan}}]{Zhang2020b}%
  \BibitemOpen
  \bibfield  {author} {\bibinfo {author} {\bibfnamefont {Y.}~\bibnamefont
  {Zhang}}, \bibinfo {author} {\bibfnamefont {Z.}~\bibnamefont {Zhan}},
  \bibinfo {author} {\bibfnamefont {F.}~\bibnamefont {Guinea}}, \bibinfo
  {author} {\bibfnamefont {J.~\'A.}\ \bibnamefont {Silva-Guill\'en}}, \ and\
  \bibinfo {author} {\bibfnamefont {S.}~\bibnamefont {Yuan}},\ }\bibfield
  {title} {\enquote {\bibinfo {title} {Tuning band gaps in twisted bilayer
  {MoS}$_2$},}\ }\href@noop {} {\  (\bibinfo {year} {2020}{\natexlab{b}})},\
  \Eprint {http://arxiv.org/abs/2005.13879} {arXiv:2005.13879
  [cond-mat.mes-hall]} \BibitemShut {NoStop}%
\end{thebibliography}%


\onecolumngrid
\newpage
\begin{center}
{\huge 
Supplemental Material 
}
\end{center}
\twocolumngrid

\renewcommand{\thetable}{S\arabic{table}}
\renewcommand{\theHtable}{S\arabic{table}}
\setcounter{table}{0}

\renewcommand{\thefigure}{S\arabic{figure}}
\renewcommand{\theHfigure}{S\arabic{figure}}
\setcounter{figure}{0}

In this Supplemental Material, we first (section \ref{Sec_appendix_struc}) 
present the commensurate moir\'e structures of twisted bilayer MoS$_2$ 
(tb-MoS$_2$) that have been used in the present work. Section 
\ref{Sec_appendix_DFT} gives some details and complementary results on our 
DFT calculations. Section \ref{Sec_appendix_TB} provides the Tight-Binding 
(TB) Slater-Koster parameters for tb-MoS$_2$. Complementary TB results 
(bands, local density of states, and eigenstates) are presented in section 
\ref{Sec_appendix_TB_results}.

\section{tb-MoS$_2$ commensurate structures}
\label{Sec_appendix_struc}

The atomic structure of commensurate twisted bilayer MoS$_2$ (tb-MoS$_2$) 
is explained in the main text. The structures of tb-MoS$_2$ that have been 
used in the present work are listed table \ref{Tab_bilayer}.

In tb-MoS$_2$, different types of moir\'e patterns can be built since the atoms of a monolayer unit cell are not equivalent by symmetry.
For our study we consider two kinds of moir\'e patterns:
\begin{itemize}
\item {\bf Patterns from AA:} Starting from an AA stacked bilayer (where Mo atoms of a layer lie above a Mo atom of the other layer, and  S atoms of a layer lie above an S atom of the other layer), the layer 2 is rotated with respect to layer 1 by the angle $\theta$ around an axis containing two Mo atoms. 
\item {\bf Patterns from AB:} Starting from an AB stacked bilayer (were Mo atoms of layer 1 lie above a Mo atom of layer 2, and S atoms of each layer do not lie above an atom of the other layer), layer 2 is rotated with respect to layer 1 by the angle $\theta$ around an axis containing two Mo atoms. 
\end{itemize}
For simplicity, in the main text we discussed only moir\'e patterns built from AA stacking, but results for moir\'e patterns built from AB stacking yield similar results, as shown here. 

\begin{table}[t!]
\caption{\label{Tab_bilayer} 
$(n,m)$ twisted bilayer MoS$_2$ (tb-MoS$_2$) structures that have been used in the present work.
$\theta$ is the rotation angle between the two layers and $N$ the number of atoms in a unit cell.
}
\begin{tabular}{lrr}
\hline \hline
{($n,m$)}~~~  &  ~~~~~~~$\theta$ [deg.]   &  ~~~~~~~~~~~~~~~~$N$    \\ \hline
(1,2)    & 21.787  &   42      \\
(2,3)    & 13.174  &  114    \\
(3,4)    &  9.430  &  222      \\
(4,5)    &  7.341  &  366  \\
(5,6)    &  6.009  &  546  \\
(6,7)    &  5.086  &  762    \\
(7,8)    &  4.408  &  1014     \\
(10,11)    &  3.150&  1986   \\
(15,16)    &  2.134  &  4326  \\
(16,17)    &  2.004  &  4902  \\
(18,19)   & 1.788   &  6162   \\
(19,20)   & 1.696   &  6846   \\
(20,21)  &  1.614  &   7566   \\
(22,23)  &  1.470  &  9114    \\
(25,26)  &  1.297 & 11706    \\
(27,28)  &  1.203 & 13614     \\
(30,31)  &  1.085 & 16746     \\
(33,34)  &  0.987 & 20202     \\
(36,37)  &  0.906 & 23982 \\
\hline \hline
\end{tabular}
\end{table}

\subsection{Moir\'e pattern from AA}
\label{sec_built_fromAA}

\begin{figure}[tb!]
\begin{center}

\includegraphics[width=1.0\columnwidth,keepaspectratio]{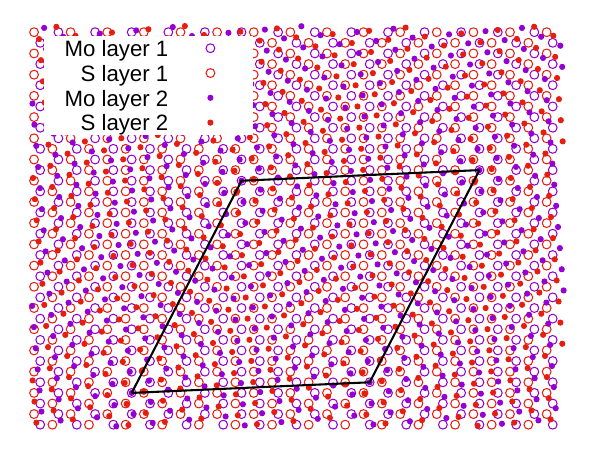}

\end{center}
\caption{\label{Fig_6-7_structure}
Atomic structure of (6,7) tb-MoS$_2$ built from AA stacked bilayers. Black lines show the unit cell. AA stacking regions are at the corners of this cell, BA' and AB' stacking regions are at $1/3$ and $2/3$ of its longest diagonal, respectively.}
\end{figure}

Figure \ref{Fig_6-7_structure} shows a top view of the atomic structure of 
$(6,7)$ tb-MoS$_2$ built from AA stacking. One can identify several 
specific types of stacking regions:

\begin{itemize}
\item AA stacking regions are regions where Mo atoms of a layer lie above a Mo atom of the other layer, and  S atoms of a layer lie above an S atom of the other layer. 
\item AB' stacking regions are regions where Mo atoms of layer 1 lie above an S atom of layer 2, and S atoms of layer 1 (Mo atoms of layer 2) do not lie above an atom of layer 2 (layer 1). 
\item BA' stacking regions are regions where S atoms of layer 1 atoms lie above a Mo atom of layer 2, and Mo atoms of layer 1 (S atoms layer 2) do not lie above an atom of layer 2 (layer 1).
\end{itemize}
In Fig.\ \ref{Fig_6-7_structure}, AA stacking regions are located at the corners of the moir\'e cell. BA' and AB' stacking regions are located at $1/3$ and $2/3$ of its long diagonal, respectively. 

\subsection{Moir\'e pattern from AB}
\label{sec_built_fromAB}

\begin{figure}[t!]
\begin{center}

\includegraphics[width=1.0\columnwidth,keepaspectratio]{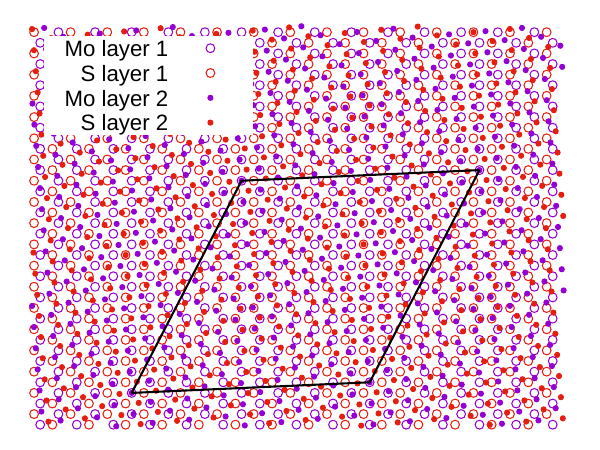}

\end{center}
\caption{\label{Fig_6-7-fAB_structure}
Atomic structure of (6,7) tb-MoS$_2$ built from AB stacked bilayers. Black lines show the unit cell. AB stacking regions are at the corners of this cell, AA' and BA stacking regions are at $1/3$ and $2/3$ of the longest diagonal, respectively.}
\end{figure}

Figure \ref{Fig_6-7-fAB_structure} shows a top view of the atomic 
structure of $(6,7)$ tb-MoS$_2$ built from AB stacking. Here one can 
identify several specific types of stacking regions:

\begin{itemize}
\item AA' stacking regions are regions where Mo atoms (S atoms) of one layer lie above an S atom (Mo atom) of the other layer.
\item AB stacking regions are regions where Mo atoms of layer 1 lie above a Mo atom of layer 2, and S atoms of each layer do not lie above an atom of the other layer.  
\item BA stacking regions are regions where S atoms of layer 1 lie above an S atom of layer 2, and Mo atoms each layer do not lie above an atom of the other layer.
\end{itemize}
In Fig.\ \ref{Fig_6-7-fAB_structure}, AB stacking regions are located at the corners of the moir\'e cell. AA' and BA stacking regions are located at $1/3$ and $2/3$ of its long diagonal, respectively.

\section{DFT tb-MoS$_2$ bands}
\label{Sec_appendix_DFT}

\begin{figure}[t!]
\includegraphics[width=1.0\columnwidth,keepaspectratio]{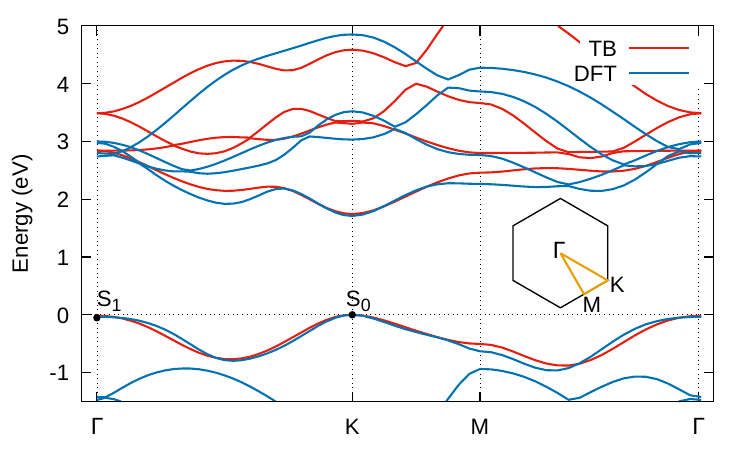}
\caption{\label{Fig_mono_DFT_TB}
DFT and TB bands in monolayer MoS$_2$. 
The origin of the energy is chosen at the maximum energy of the valence band, {\it i.e.}, at the energy of the states labeled S$_0$, $E({\rm S}_0)=0$. Since $E({\rm S}_1) < E({\rm S}_0)$ the gap is direct at K. The first Brillouin zone is sketched in the insert. 
}
\end{figure}

Density functional theory (DFT) \cite{Hohenberg64,Kohn65} calculations 
based on first-principle calculations were carried out with the ABINIT 
code \cite{Gonze02,Gonze09,Gonze16}, using the local density approximation 
(LDA) exchange-correlation functional \cite{Jones89} and the 
Perdew-Burke-Ernzerhof (PBE) parametrized generalized gradient 
approximation (GGA) exchange-correlation functional \cite{Perdew96}. We 
considered fourteen valence electrons of Mo $(4s^2,4p^6,4d^5,5s^1)$, six 
valence electrons of S $(3s^2,3p^4)$ in the PAW-PBE pseudopotential. The 
Brillouin zone was sampled by a k-point mesh of 0.8\,nm$^{-1}$ separation 
in reciprocal space within the Monkhorst-Pack scheme \cite{Zupan98}, and 
the kinetic energy cutoff was chosen to be 544.22\,eV. A vacuum region of 
2\,nm was inserted between the MoS$_2$ bilayers to avoid spurious 
interactions between periodic images.

Figure \ref{Fig_mono_DFT_TB} shows our DFT results for the band structure 
of monolayer MoS$_2$. The highest energy of the valence band (state S$_0$ 
at the point K) is fixed to zero, $E({\rm S}_0)=0$. Since the highest 
valence energy at the point $\Gamma$ (state S$_1$) has a lower energy, the 
gap is direct, as expected \cite{Huisman71}.

\begin{figure}[t!]
\begin{center}

\includegraphics[width=1.0\columnwidth,keepaspectratio]{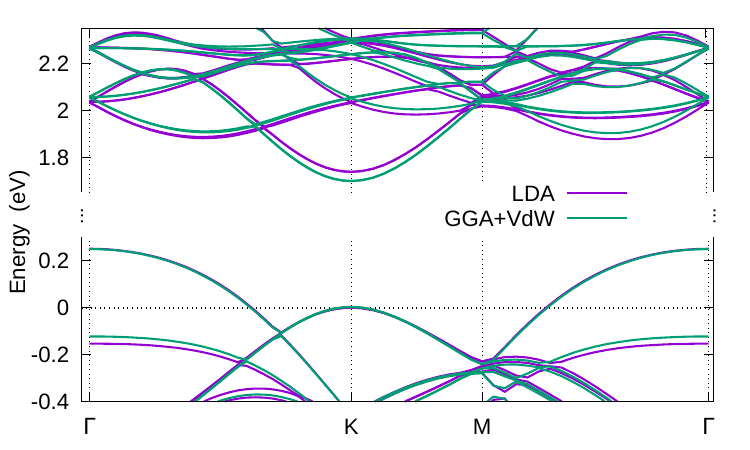}

\end{center}
\caption{\label{Fig_DFT_LDA_GGA-VdW}
DFT bands around the gap in $(1,2)$ $\theta = 21.79^\circ$ tb-MoS$_2$ 
(built from AA stacking, see Sec.\ \ref{sec_built_fromAA}): Comparison 
between LDA and the PBE-GGA + Van der Waals exchange-correlation 
functional.
}
\end{figure}

Figure \ref{Fig_DFT_LDA_GGA-VdW} shows that LDA and GGA + Van der Waals 
approximations yield very similar results, so all further result are based 
on LDA calculations.

\begin{figure}[t!]
\begin{center}
\includegraphics[width=1.0\columnwidth,keepaspectratio]{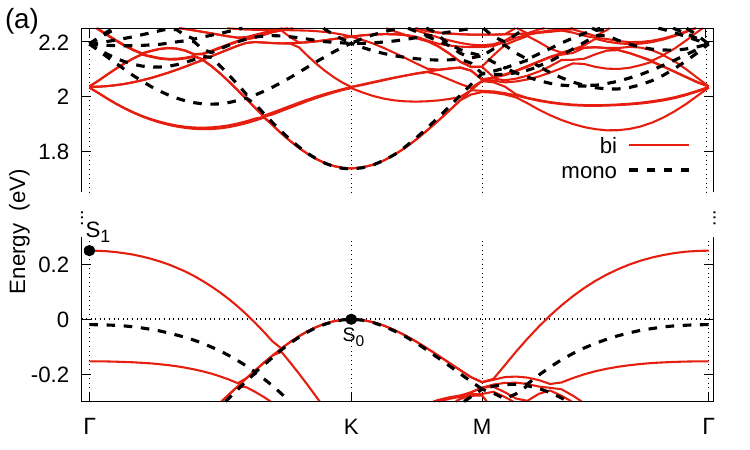}

\includegraphics[width=1.0\columnwidth,keepaspectratio]{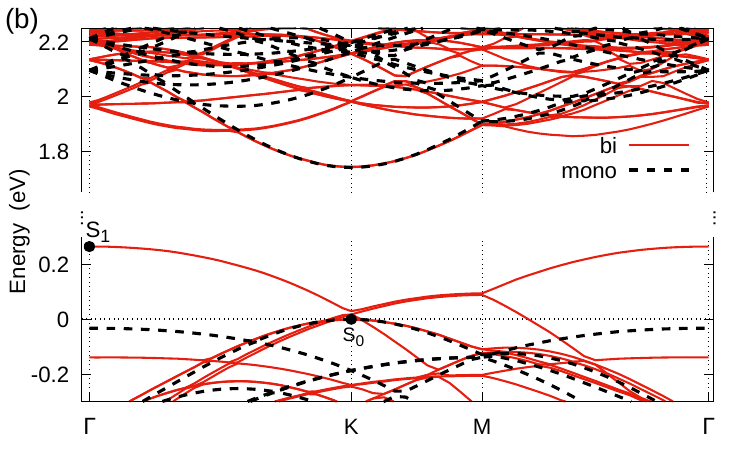}

\end{center}
\caption{\label{Fig_DFT_bi_mono}
DFT bands around the gap in tb-MoS$_2$  (built from AA stacking, see Sec.\ \ref{sec_built_fromAA}): 
(a) $(1,2)$ $\theta = 21.79^\circ$, 
(b) $(2,3)$ $\theta = 13.17^\circ$.
(red color) bilayer, (black color) monolayer represented in the bilayer unit cell.
}
\end{figure}

In Fig.\ \ref{Fig_DFT_bi_mono}, we compare the bands around the gap of 
$(1,2)$ and $(2,3)$ tb-MoS$_2$ with the monolayer case. For the purpose of 
comparison, the monolayer unit cell has been mapped to the bilayer one.

\section{Tight-Binding (TB) model}
\label{Sec_appendix_TB}

In this section, we present details of our Tight-Binding (TB) Hamiltonian 
for monolayer MoS$_2$ and twisted bilayer MoS$_2$. We use a Slater-Koster 
scheme \cite{Slater54} in order to describe all the studied structures by 
a single set of parameters.

\subsection{TB model for monolayer MoS$_2$}

We start by describing a single layer of 2H-MoS$_2$. The lattice vectors 
of monolayer MoS$_2$ are $\vec{a}_1$ ($a\sqrt{3}/2,-a/2,0)$ and 
$\vec{a}_2$ ($a\sqrt{3}/2,a/2,0)$, with the lattice parameter 
$a=0.318$\,nm. A unit cell contains 3 atoms: Mo at $(0,0,0)$, S at 
$(0,a/\sqrt{3},0.49115 a)$ and S at $(0,a/\sqrt{3},-0.49115 a)$ 
\cite{Huisman71,Ridolfi15}.

Our TB model includes 11 orbitals per unit cell of the monolayer: 5 $d$ Mo 
orbitals ($d_0 = 4d_{z^2}$, $d_1 = 4d_{xz},\, 4d_{yz}$, $d_2 = 
4d_{x^2-y^2},\,4d_{xy}$ of 1 Mo atom) and 6 $p$ S orbitals ($3p_x$, $3p_y$ 
and $3p_z$ of 2 S atoms). By symmetry, the 2 $d_1$ ($d_2$) orbitals of 
each Mo are equivalent as well as the $3p_x$ and $3p_y$ of each S.

\begin{table}[t!]
\caption{\label{Tab_TB_mono} 
Tight-binding (TB) Slater-Koster parameters for monolayer MoS$_2$, and pairs of neighbors for which the hopping term is nonzero. The lattice parameter of monolayer MoS$_2$ is $a=0.318$\,nm.
}
\begin{tabular}{l|l|l}
\hline \hline
Atom~~~~~~~~~~~~~~~~~~ & Orbitals~~~~~~~~~~~~~~~~~~~~~~~~~~~~~~ & On-site energy (eV) \\ \hline
Mo     & $d_0 = 4d_{z^2}$  & $E^0_0 = 0.1356 $  \\
       & $d_1 = 4d_{xz},\, 4d_{yz}$  & $E^0_1 = -0.4204 $  \\
       & $d_2 = 4d_{x^2-y^2},\,4d_{xy}$  & $E^0_2 = 0.0149$  \\ \hline
S      & $3p_x$,  $3p_y$ & $E^0_{x,y} =-38.71 $ \\
       & $3p_z$          & $E^0_{z} = -29.45 $ \\ 
\end{tabular}
\begin{tabular}{l|l|l|l|l}
\hline \hline
Atom~~~ & Neighbor~~~ & Number ~~~& Inter-atomic ~~~ & Slater-Koster~~~~~~~~~ \\
& &  &distance (nm) & parameters (eV)   \\ \hline
Mo & Mo  & 3    & 0.318     &  $V_{dd\sigma} = -0.9035 $   \\
   &      &      &           &  $V_{dd\pi} = 0.7027$       \\ 
   &      &      &           &  $V_{dd\delta} = 0.0897$  \\ \cline{2-5}
 & S & 6    & 0.241     &  $V_{dp\sigma} = -7.193 $   \\
 &   &      &           &  $V_{dp\pi} = 3.267 $       \\ \hline
S & S & 1    & 0.312     &  $V_{pp\sigma} = 8.079 $   \\
  &    &      &           &  $V_{pp\pi} = -2.678 $       \\ \cline{3-5}
  &    & 6    & 0.318     &  $V_{pp\sigma} = 7.336$   \\
  &    &      &           &  $V_{pp\pi} = -2.432 $       \\ 
\hline \hline
\end{tabular}
\end{table}

Our TB model for monolayer MoS$_2$ is an adaptation of the model proposed 
in Ref.~\cite{Ridolfi15} to our DFT results (Fig.\ \ref{Fig_mono_DFT_TB}).  
The $p\,{\rm S}-p\,{\rm S}$ , $d\,{\rm Mo}-d\,{\rm Mo}$ and $d\,{\rm 
Mo}-p\,{\rm S}$ hopping terms are calculated using a Slater-Koster formula 
with the parameters $V_{pp\sigma}$, $V_{pp\pi}$, $V_{dd\sigma}$, 
$V_{dd\pi}$, $V_{dd\delta}$, $V_{dp\sigma}$, $V_{dp\pi}$. For the 
monolayer, only first neighbor S-S, Mo-Mo and S-Mo hopping terms are taken 
into account. On-site energy values, number of neighbors taken into 
account, and values of Slater-Koster parameters are listed table 
\ref{Tab_TB_mono}.

\subsection{TB model of twisted bilayer MoS$_2$ (tb-MoS$_2$)}

\begin{table}[tb!]
\caption{\label{Tab_TB_interlayer} 
TB Slater-Koster parameters for interlayer hopping terms in tb-MoS$_2$. $d^0$ is the interlayer distance; for the definition of $d_{\rm Mo-Mo}$, $d_{\rm Mo-S}$, and $d_{\rm S-S}$, see Fig.\ \ref{Fig_structure2-3}.
}
\begin{tabular}{l|l|l|l|l}
\hline \hline
Atom~~~~~ & Neighbor~~~~~ & $d^0$ (nm) ~~~~~  & $q$ ~~~~~~~~~~~~~~ & Slater-Koster~~~~~~~~~ \\
& & & & parameters (eV)   \\ \hline
Mo & Mo  & 0.6800  & 11.6496  &  $V^0_{dd\sigma} = -0.1416  $   \\
   &     &       &   &  $V^0_{dd\pi} = -0.4254 $   \\
   &     &       &   &  $V^0_{dd\delta} =  -0.1237 $   \\ \cline{2-5}
   & S   & 0.5238  & 8.9738  &  $V^0_{dp\sigma} = -1.4793 $   \\
   &     &       &   &  $V^0_{dp\pi} = 0.52431 $   \\ \hline
S  & S   & 0.3676  & 6.2981  &  $V^0_{pp\sigma} =  6.2782 $   \\
   &     &       &   &  $V^0_{pp\pi} = -8.9733 $   \\
\hline \hline
\end{tabular}
\end{table}

Now we move to tb-MoS$_2$ where we take the monolayer MoS$_2$ hopping 
terms of table \ref{Tab_TB_mono} as the intralayer hopping terms. Most 
previous studies \cite{Cappelluti13, Roldan14b,Fang15,Zahid13} include 
only $p\,{\rm S}-p\,{\rm S}$ interlayer hopping terms. At first sight, 
this may appear justified since these correspond to the shortest 
interlayer distance, see Fig.\ \ref{Fig_structure2-3}. However, $d_{{\rm 
Mo}-{\rm Mo}}$ and in particular $d_{{\rm Mo}-{\rm S}}$ are not that much 
bigger than $d_{{\rm S}-{\rm S}}$ such that a sharp cutoff at the shortest 
distance may not be appropriate. Indeed, $d\,{\rm Mo}-p\,{\rm S}$ terms 
and $d\,{\rm Mo}-d\,{\rm Mo}$ terms may also be important because we do 
not limit the interlayer coupling to first-neighbor hopping. Therefore, we 
include $p\,{\rm S}-p\,{\rm S}$, $d\,{\rm Mo}-p\,{\rm S}$, and $d\,{\rm 
Mo}-d\,{\rm Mo}$ interlayer hopping terms in our Slater-Koster scheme. 
Following other studies of twisted bilayer TMDCs \cite{Fang15} and twisted 
bilayer graphene \cite{Trambly10,Trambly12}, each interlayer Slater-Koster 
parameter $V_i$ is assumed to decrease exponentially as a function of the 
distance $d$ between orbitals:
\begin{equation}
\label{Eq_Interlayer_SK}
V_i(d) = V^0_i \exp \left( - q_i \frac{d-d_0}{d_0} \right) \, F_c(d) \,, 
\end{equation}
where the $V^0_i$ is S-S $V^0_{pp\sigma}$, S-S $V^0_{pp\pi}$, Mo-Mo 
$V^0_{dd\sigma}$, Mo-Mo $V^0_{dd\pi}$, Mo-Mo $V^0_{dd\delta}$, Mo-S 
$V^0_{dp\sigma}$, Mo-S $V^0_{dp\pi}$, respectively; $d^0_i$ is the 
corresponding interlayer distance $d_{\rm S-S}$, $d_{\rm Mo-Mo}$, and 
$d_{\rm Mo-S}$, respectively (see Fig.\ \ref{Fig_structure2-3}). The 
coefficients $q_i$ are fixed, like in twisted bilayer graphene 
\cite{Trambly12}, to have a reduction by a factor 10 between first 
neighbor hopping and second neighbor hopping terms,
\begin{equation}
q_i = \frac{\sqrt{3}\, \ln (10) \, d^0_i}{(\sqrt{3}-1)\, a}.
\end{equation}
Numerical values of $V^0_i$, $d^0_i$ and $q_i$ are listed table 
\ref{Tab_TB_interlayer}.

In equation (\ref{Eq_Interlayer_SK}) a smooth cutoff function 
\cite{Mehl96} is used,
\begin{equation}
F_c(d) ~=~  \left( 1 + \exp \left( \frac{d - r_c}{l_c} \right)  \right)^{-1},
\label{Eq_cutoffFunction}
\end{equation}
with $r_c$ the cutoff distance and $l_c = 0.0265$\,nm \cite{Mehl96}. For 
$r \ll r_c$, $F_c(r) \simeq 1$; and for $r \gg r_c$, $F_c(r) \simeq 0$. 
All results presented in the present Rapid Communication are calculated
with $r_c = 2.5 a = 0.795$\,nm.

\section{TB electronic structure of tb-MoS$_2$}
\label{Sec_appendix_TB_results}

\begin{figure}[tb!]
\includegraphics[width=1.0\columnwidth,keepaspectratio]{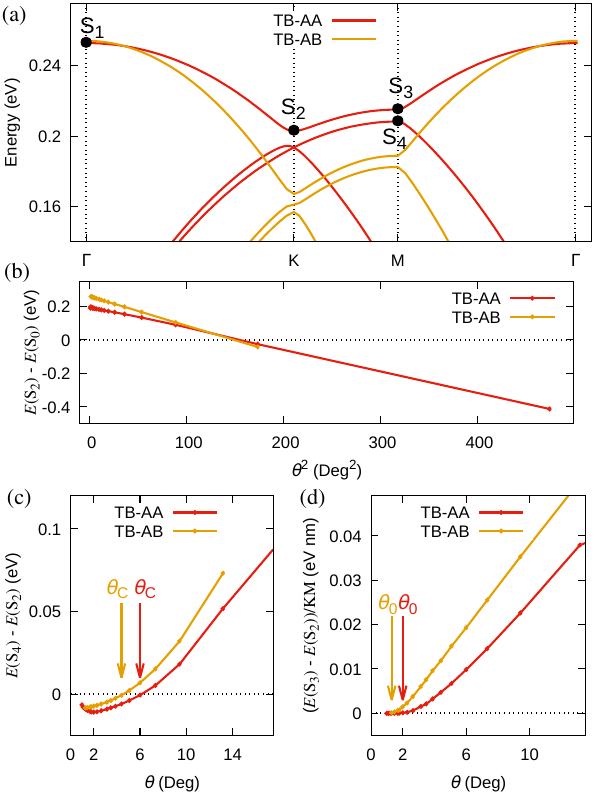}

\caption{\label{fig_analyse_Ei_AB-AA}
Dependence of valence bands on rotation angle $\theta$: Comparison between 
tb-MoS$_2$ built from AA stacking (TB-AA, see section 
\ref{sec_built_fromAA}) and tb-Mos$_2$ built from AB stacking (TB-AB, see 
section \ref{sec_built_fromAB}). The TB-AA curves coincide with those of 
Fig.\ \ref{fig_analyse_Ei}. (a) valence-band dispersion of $(4,5)$ 
tb-MoS$_2$, $\theta = 7.34^\circ$. (b) Energy $E({\rm S}_2)$ of state 
S$_2$ (see panel (a)) versus $\theta^2$. (c) Energy difference between 
states S$_4$ and S$_2$, $\Delta E_{24} = E({\rm S}_4) - E({\rm S}_2)$, 
versus $\theta$. A negative $\Delta E_{24}$ value means that a gap $ | 
\Delta E_{24} | $ exists between the band below the gap and the other 
valence bands. (d) Average slope of $E(\vec k)$ of the band between states 
S$_2$ and S$_3$. For every rotation angle, the origin of energy is fixed 
at the energy $E({\rm S}_0)$ of the state S$_0$ (see Figs.\ 
\ref{Fig_DFT_4angles} and \ref{Fig_DFT_bi_mono}).
}
\end{figure}

\begin{figure}[tb!]
\begin{center}
\includegraphics[width=1.0\columnwidth,keepaspectratio]{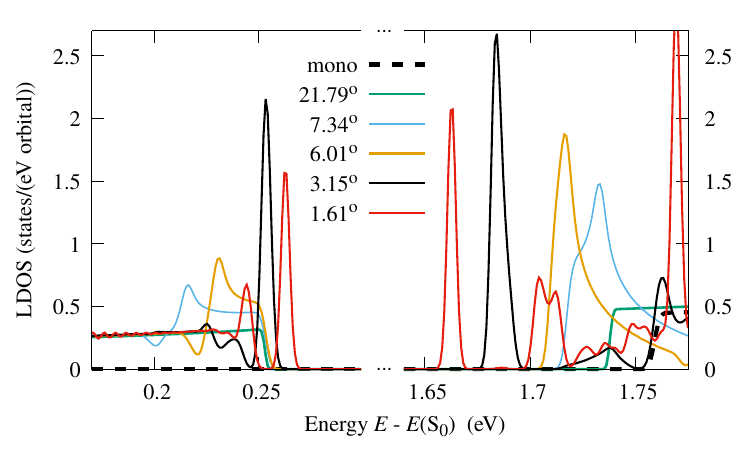}

\caption{\label{fig_LDOSAA}
TB local density of states (LDOS) of the $4d_{z^2}$ Mo orbital at the 
center of the AA stacking region for different rotation angles of 
tb-MoS$_2$. The LDOS is calculated employing a Gaussian broadening with a 
standard deviation $\sigma = 2$\,meV.
}
\end{center}
\end{figure}

\subsection{Analysis of the bands in tb-MoS$_2$ built from AB stacking}

In the main text we showed only results for tb-MoS$_2$ built from AA 
stacking (see section \ref{sec_built_fromAA}). However, other types of 
moir\'e patterns exist in this system as well; in particular, one can 
start from AB stacking (see section \ref{sec_built_fromAB}). Figure 
\ref{fig_analyse_Ei_AB-AA} presents a comparison of the 
$\theta$-dependence of the band structure between tb-MoS$_2$ between 
bilayers built from AA stacking and from AB stacking. 
The results are qualitatively very similar, which 
shows that the main results of our study do not depend on the type of 
moir\'e pattern. The main quantitative differences with respect to the 
results discussed in the main text are the values of $\theta_C$ and 
$\theta_0$: $\theta_C \approx 6^\circ$ versus $\theta_C \approx 4.5 
^\circ$ for tb-MoS$_2$ built from AA and AB stacking, respectively; 
$\theta_0 \approx 2^\circ$ versus $\theta_0 \approx 1.8^\circ$ for 
tb-MoS$_2$ built from AA and AB stacking, respectively.

\begin{figure}[tb!]
\begin{center}
\includegraphics[width=1.0\columnwidth,keepaspectratio]{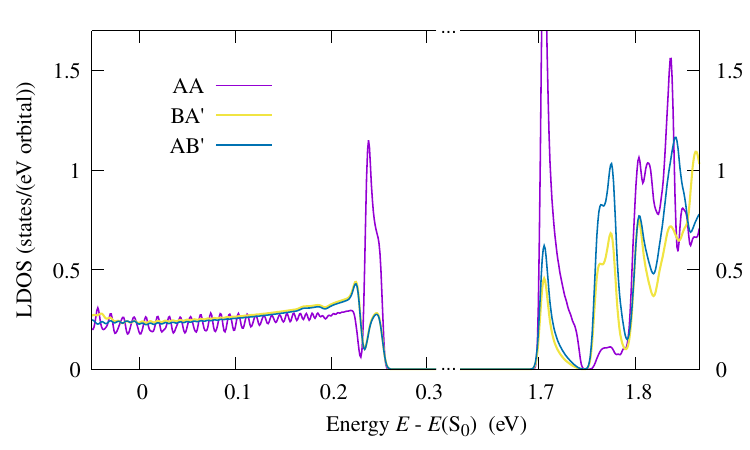}

\includegraphics[width=1.0\columnwidth,keepaspectratio]{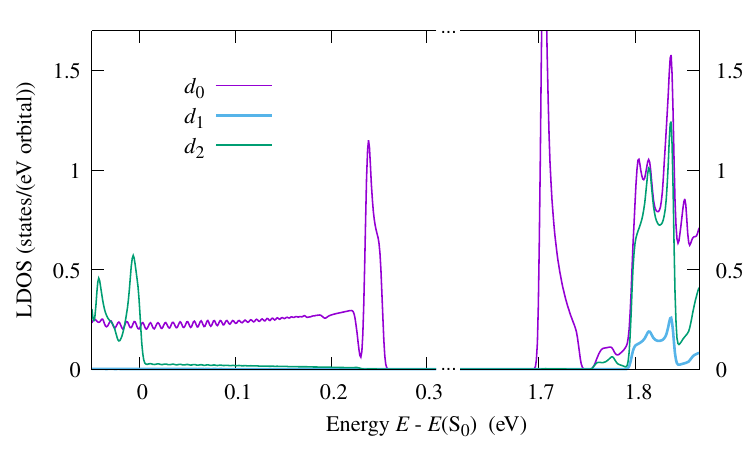}

\caption{\label{fig_LDOSAA-AB-BA}
TB local density of states (LDOS) of the Mo orbital around the main gap in 
(6,7) $\theta = 5.09^\circ$ tb-MoS$_2$ (built from AA stacking, see Sec.\ 
\ref{sec_built_fromAA}): (Top panel) LDOS of the $d_0 = d_{Z^2}$ Mo 
orbital at the center of AA, BA', and AB' stacking regions (Sec.\ 
\ref{sec_built_fromAA}). (Bottom panel) LDOS of $d_0$, $d_1 = 4d_{xz},\, 
4d_{yz}$ and $d_2 = 4d_{x^2-y^2},\,4d_{xy}$ Mo orbitals at the center of 
the AA stacking region. The LDOS is calculated employing a Gaussian 
broadening with the standard deviation $\sigma = 2$\,meV.
}
\end{center}
\end{figure}

\subsection{Local density of states (LDOS)}

The TB density of states (DOS) is calculated employing a Gaussian 
broadening with a standard deviation $\sigma = 2$\,meV. For the 
$k$-integration we use a grid with $Nk_x \times Nk_y$ points in the 
reciprocal unit cell, with $Nk_x = Nk_y$ large enough to obtain a DOS that 
is independent of these parameters. Due to this broadening, the minigaps 
found in the band structure are not always seen clearly in the DOS.

Figure \ref{fig_LDOSAA} shows the LDOS for the $d_0 = d_{z^2}$ orbital of 
an Mo atom at the center of the AA stacking region for several rotation 
angles $\theta$. Figure \ref{fig_LDOSAA-AB-BA} (top panel) shows the local 
density of states (LDOS) of the $d_0 = d_{z^2}$ Mo orbital for the 
selected rotation angle $\theta = 5.09^\circ$, but for Mo atoms located at 
different stacking regions of the moir\'e pattern (see section 
\ref{Sec_appendix_struc}). Confined states (``flat bands'') lead to sharp 
peaks in the LDOS (Fig.\ \ref{fig_LDOSAA}). These states have $d_{z^2}$ Mo 
character (Fig.\ \ref{fig_LDOSAA-AB-BA} (bottom panel)), and a very small 
weight for the other $d$ Mo orbitals. The flat bands are mainly located in 
the AA stacking region (Figs.\ \ref{fig_LDOSAA-AB-BA} (top panel) and 
\ref{Fig_20-21_VecP_G-K-M_ValenceB_average}). Figure 
\ref{Fig_20-21_VecP_G-K-M_ValenceB_average} shows that the lowest-energy 
flat bands in the conduction and valence bands correspond to Mo atoms that 
are located at the center of the AA stacking regions, and that the next 
flat band in the conduction and valence bands corresponds to states 
located in a ring in the AA stacking regions.

\subsection{Eigenstates corresponding to flat bands in tb-MoS$_2$}

\begin{figure*}[p!]
\begin{center}
\includegraphics[width=1.99\columnwidth,keepaspectratio]{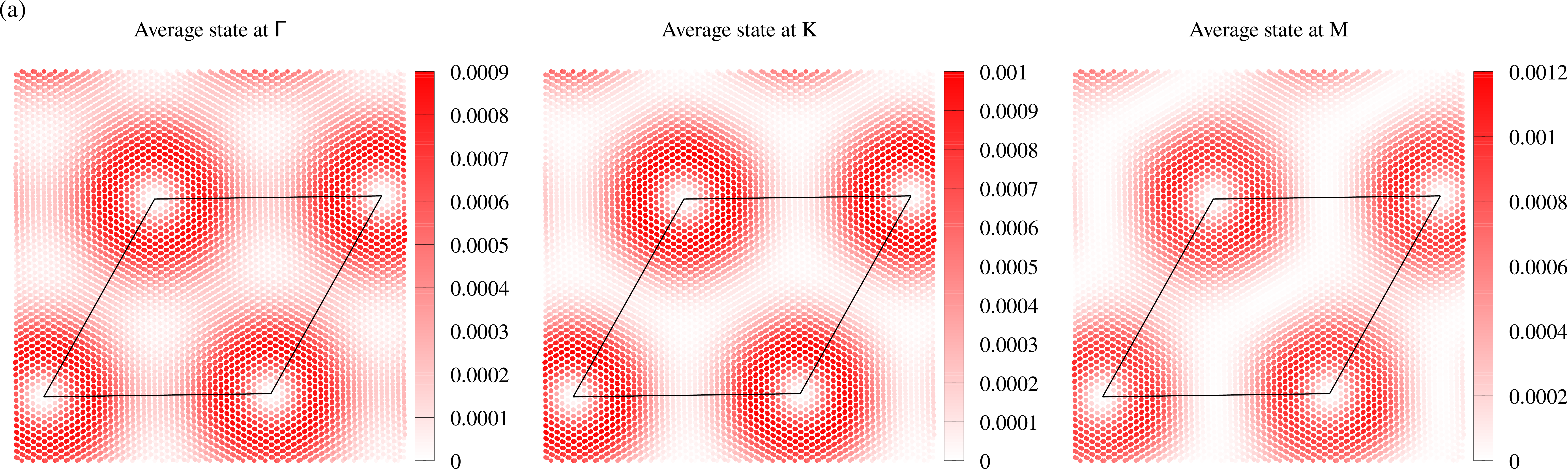} \\[2mm]
\includegraphics[width=1.99\columnwidth,keepaspectratio]{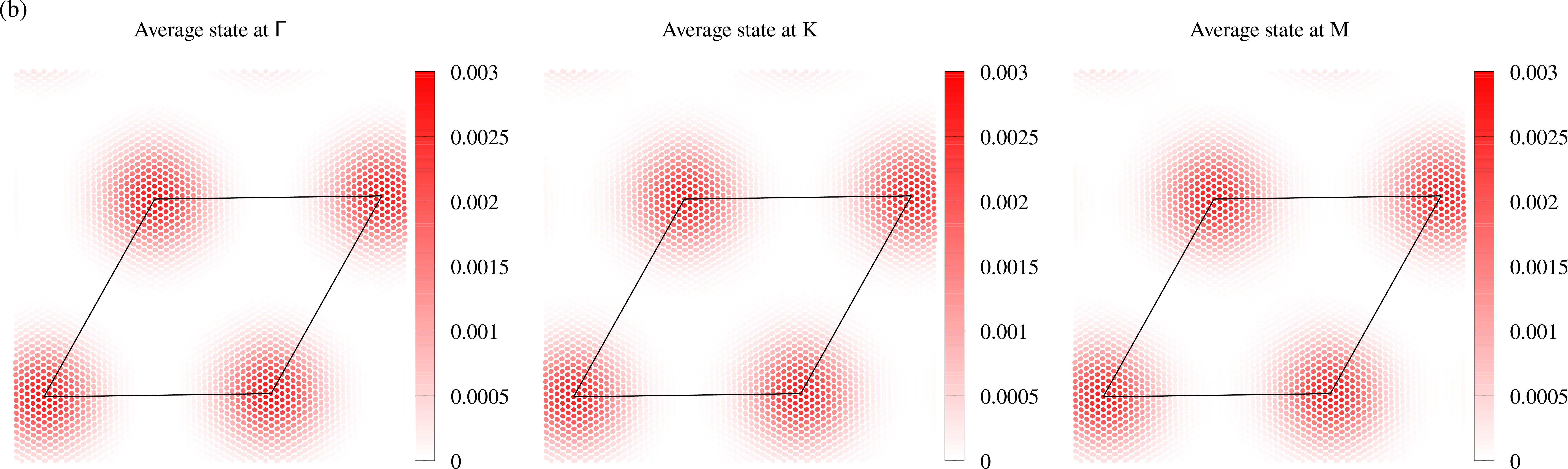} \\[2mm]
\includegraphics[width=1.99\columnwidth,keepaspectratio]{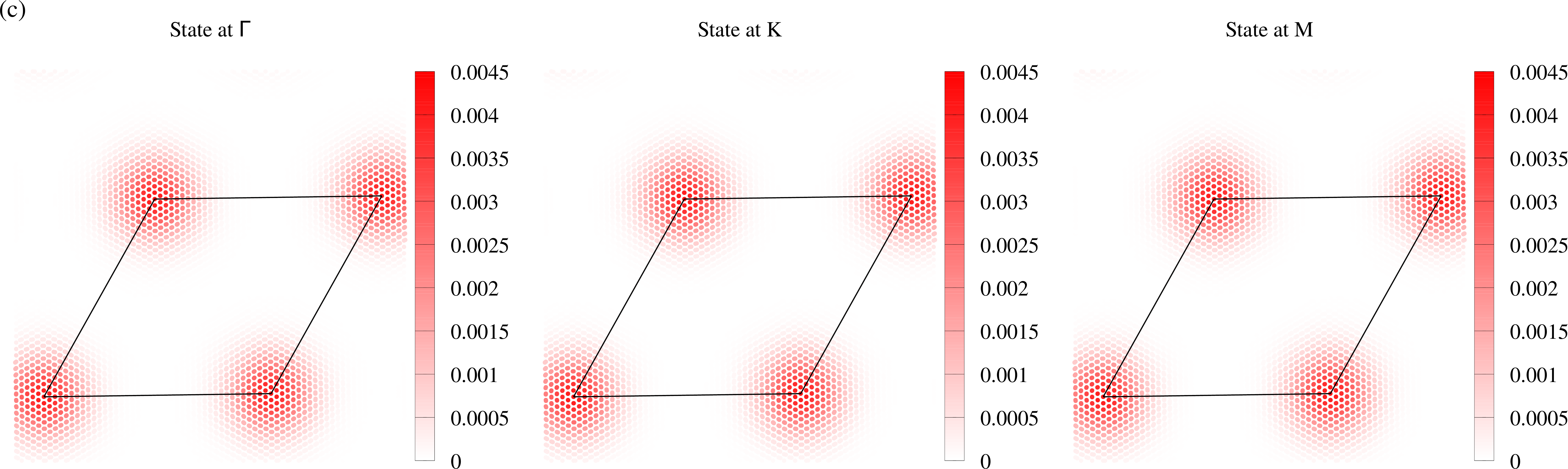} \\[2mm]
\includegraphics[width=1.99\columnwidth,keepaspectratio]{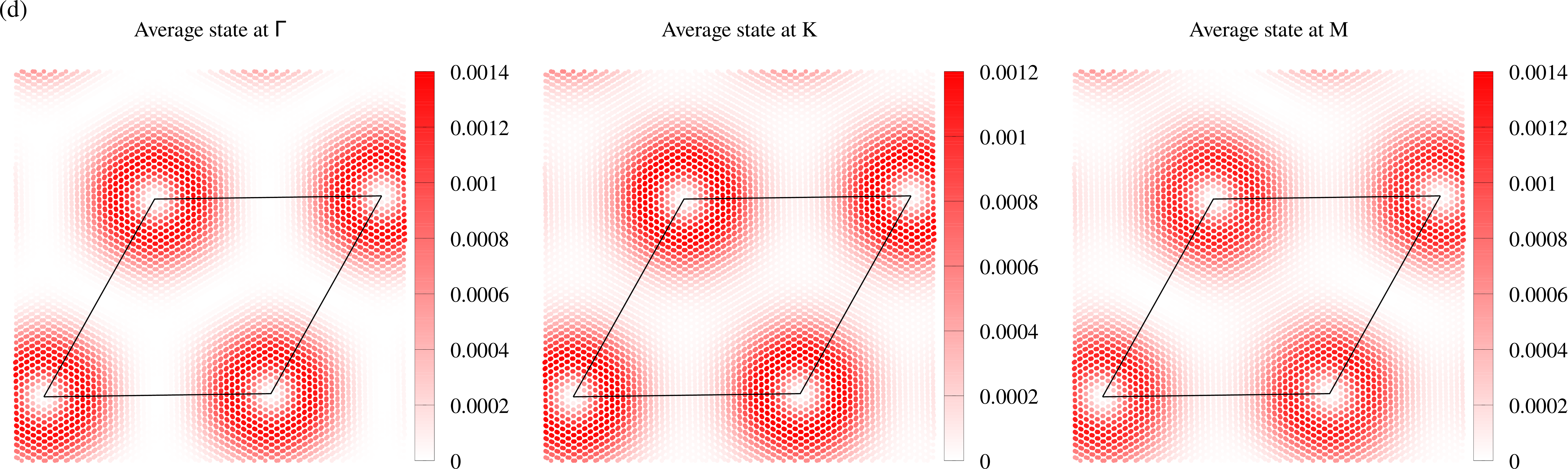}
\end{center}
\caption{\label{Fig_20-21_VecP_G-K-M_ValenceB_average}
Average weight of the eigenstates at $\Gamma$, K, and M, of the flat band around the gap in real space in $(20,21)$ tb-MoS$_2$ (built from AA stacking, see Sec.\ \ref{sec_built_fromAA}) with a rotation angle  $\theta = 1.61^\circ$:
Conduction band: (a) Average of the four-fold quasi-degenerate band at energy 
$E \simeq E({\rm S}_0) + 1.686 \pm 0.002$\,eV 
and (b) average of the two-fold quasi-degenerate band at energy 
$E \simeq E({\rm S}_0) + 1.6626 \pm 0.0002$\,eV.
Valance band: (c) non-degenerate bands at energy 
$E \simeq E({\rm S}_0) + 0.26249 \pm 0.00001$\,eV,
and (d) average of the two-fold quasi-degenerate band at energy
$E \simeq E({\rm S}_0) + 0.2518 \pm 0.0003$\,eV.
The corresponding bands are shown in Fig.\ \ref{fig_LDOSAA_bnds}(b). The 
color scale shows the weight of the eigenstate on each $d_0 = 4d_{z^2}$ 
orbital of the Mo atoms. The sum of these weights is more than 98\% and 
95\% of each state for the valence and conduction band, respectively. 
Black lines show the unit cell containing 2522 Mo atoms. AA stacking 
regions are at the corners of this cell.
}
\end{figure*}

Analysis of the band dispersion (Fig.\ \ref{fig_LDOSAA_bnds}) shows that 
the first isolated flat band below the main gap (valence band) is 
non-degenerate and thus contains one state per moir\'e cell. By contrast, 
the two first isolated flat bands above the gap (conduction bands) are 
two-fold quasi degenerate.  The weight of the eigenstates corresponding to 
these flat bands is mainly concentrated on $d_0 = d_{z^2}$ Mo orbitals 
(more than 98\% and 95\%, respectively) located at the center of AA 
stacking regions. This is shown in 
Fig.~\ref{Fig_20-21_VecP_G-K-M_ValenceB_average}(b,c) for eigenstates at 
the points $\Gamma$, K, and M of the flat band above and below the main 
gap, respectively.

For small enough angles, the next isolated flat bands 
(Fig.~\ref{fig_LDOSAA_bnds}(b)) are four-fold quasi degenerate in the 
conduction band and two-fold quasi degenerate in the valence band. The 
weight of the eigenstates, corresponding to these flat bands at the points 
$\Gamma$, K, and M, is mainly located in a ring in AA stacking regions 
(Fig.~\ref{Fig_20-21_VecP_G-K-M_ValenceB_average}(a,d)).

\newpage

\end{document}